\documentclass[pdflatex,sn-mathphys-num]{sn-jnl}

\usepackage{hhline}
\usepackage{graphicx}%
\usepackage{multirow}%
\usepackage{amsmath,amssymb,amsfonts}%
\usepackage{amsthm}%
\usepackage{mathrsfs}%
\usepackage[title]{appendix}%
\usepackage{xcolor}%
\usepackage{textcomp}%
\usepackage{manyfoot}%
\usepackage{booktabs}%
\usepackage{algorithm}%
\usepackage{algorithmicx}%
\usepackage{algpseudocode}%
\usepackage{listings}%
\usepackage{todonotes}
\usepackage{multirow}
\usepackage{cleveref}

\theoremstyle{thmstyleone}%
\theoremstyle{thmstyletwo}%

\theoremstyle{thmstylethree}%

\begin{document}

\title[Article Title]{A Machine Learning Framework for Constructing Heterogeneous Contact Networks: Implications for Epidemic Modelling}


\author*[1,2]{\fnm{Luke} \sur{Murray Kearney}}\email{luke.murray-kearney@warwick.ac.uk}

\author[2,3]{\fnm{Emma L.} \sur{Davis}}\email{emma.l.davis@warwick.ac.uk}
\equalcont{These authors contributed equally to this work.\newpage}

\author[2,4]{\fnm{Matt J.} \sur{Keeling}}\email{m.j.keeling@warwick.ac.uk}
\equalcont{These authors contributed equally to this work.\newpage}

\affil*[1]{\orgdiv{MathSys Centre for Doctoral Training}, \orgname{Mathematics Institute, and School of Life Sciences}, \orgaddress{\street{University of Warwick}, \city{Coventry}, \country{UK}}}

\affil[2]{\orgdiv{Zeeman Institute for Systems Biology and Infectious Disease Epidemiology Research (SBIDER)}, \orgname{University of Warwick}, \orgaddress{\city{Coventry}, \country{UK}}}

\affil[3]{\orgdiv{Statistics Department}, \orgname{University of Warwick}, \orgaddress{\city{Coventry}, \country{UK}}}

\affil[4]{\orgdiv{Mathematics Institute and School of Life Sciences}, \orgname{University of Warwick}, \orgaddress{\city{Coventry}, \country{UK}}}


\abstract{
Capturing the structured mixing within a population is key to the reliable projection of infectious disease dynamics and hence informed control. Both heterogeneity in the number of epidemiologically-relevant contacts and age-structured mixing have been repeatedly demonstrated as fundamental, yet are rarely combined. Networks provide a powerful and intuitive method to realise these two elements of population structure, and simulate infection dynamics.  While there are a few key examples of contact networks being measured explicitly, this is not scalable to larger populations, where representative networks must be constructed from more ubiquitous individual-level data.
Here, using data from social contact surveys, we develop a generalisable and robust algorithm utilizing machine learning to generate a surrogate population-scale network that preserves both age-structured mixing and heterogeneity of contacts. 
For different datasets and network construction assumptions we simulate the spread of infection, considering how the epidemic size varies over basic reproduction number ($R_0$) scenarios - mirroring the process of determining public health impact from early epidemic growth.
Our approach shows that both age structure and degree heterogeneity substantially reduce the epidemic size (for a given $R_0$) compared to simpler models. We also demonstrate that these simulations more accurately re-capture the heterogeneity in secondary cases that has been observed, when transmission is scaled by contact duration to dampen the effect of highly connected nodes (``super-spreaders").
By using survey data collected during 2020-2022, these network models also inform about the impacts of control and targeting of public health interventions: quantifying the non-linear reduction in transmission opportunities that occurred during lockdowns, and the ages and contact types most responsible for onward transmission. 
Our robust methodology therefore allows for the inclusion of the full wealth of data commonly collected by surveys but frequently overlooked to be incorporated into more realistic transmission models of infectious diseases.
}



\maketitle

\section*{Introduction}\label{sec: intro}
Mathematical modelling of infectious diseases has become an integral process in shaping both public health response measures to epidemics and pandemic preparedness~\cite{brooks2021modelling,funk2020short, biggerstaff2016results, viboud2018rapidd}. Historically, early epidemiological models assumed that the population of interest was homogeneous and `well-mixed'. Under these assumptions, all infectious individuals have an equal rate of transmission to any susceptible individual in the population; leading to the determinstic $SIR$ equations. While this is an over-simplification, it has provided a robust and surprisingly accurate method of predicting infection dynamics and guiding public health decisions  \cite{keeling2008modeling,dashtbali2021compartmental, molla2023mathematical, zhan2019real, salem2016mathematical}. In reality, transmission is frequently linked to proximity and social contacts, as exemplified by the commonly-used risk thresholds for COVID-19 transmission (being within 2m for 15 minutes)~\cite{ferretti2024digital}. Based on pioneering work from the social sciences~\cite{klovdahl1994social}, there has been a growing interest in capturing patterns of human social contacts and the networks that are implied~\cite{eubank2004modelling, eames2015six} to inform infectious disease models.

When the links of a network represent routes for possible transmission (sexual contacts, close social interactions, or spatial proximity dependent on the given pathogen) then this network embeds much of the important epidemiological information. In particular, the heterogeneity in network contacts (referred to as network degree) is linked to the heterogeneities in secondary case distribution recorded for many infections~\cite{galvani2005dimensions, lloyd2005superspreading, adam2020clustering, de2013largest, shen2004superspreading}. 
Networks can also capture other structures such as: assortative mixing which amplifies the role of superspreaders~\cite{garnett1993contact, britton2024improving}; clustering which enhances local transmission but reduces wider dissemination~\cite{keeling1999effects, volz2011effects}; and long-range contacts which interconnect entire populations promoting rapid spread of infections ~\cite{watts1998collective, may2006network,danon2011networks}.

However, in practice, obtaining the complete information needed to define a population-level contact network is extremely challenging, although several attempts have been made. The use of electronic devices (wearable RFID sensors~\cite{salathe2010high,kiti2016quantifying} or Bluetooth enabled smartphones~\cite{leith2020coronavirus}) provides an efficient method of data capture, but only informs about connections {\it within} the participating population. Contact tracing data can also generate networks of transmission events~\cite{kleinman2020digital,ferretti2024digital} but these often describe only realised transmission routes, may miss unknown or casual contacts, and are biased towards individuals who have already been infected. 

Instead, much of our knowledge of human contact patterns comes from respondent-based surveys, following the foundational NATSAL~\cite{johnson1992sexual, mitchell2013sexual} and POLYMOD~\cite{mossong2008social} studies. These surveys record contacts reported by individual respondents, but generally do not observe how those contacts connect to each other within the wider population. Such surveys have been refined over time~\cite{danon2012social, gimma2022changes} and now provide a key component of epidemiological models; they have been collated into open source platforms like \href{https://socialcontactdata.org/}{socialcontactdata.org}, providing a standardized syntax. In general, these respondent-driven surveys capture both degree heterogeneity and age-structured mixing.

Despite this, many epidemiological modelling studies ignore the individual-level heterogeneity of contacts in favour of more general average patterns. Most commonly, age-structured mixing matrices are used to capture the average level of contact between age groups~\cite{mossong2008social}. While this has provided the foundation for many important epidemiological studies~\cite{davies2020age,bedford2015global}, it neglects the observable heterogeneity in social contacts, despite this heterogeneity's historical importance for sexually transmitted infections~\cite{may1987commentary, anderson1986preliminary}.

In this study, we formulate a novel method for the accurate reconstruction of age-structured networks from commonly collected survey data, which preserves both age-dependent mixing and contact heterogeneity. We demonstrate the potential of this methodology using snapshots of the CoMix survey data \cite{gimma2022changes} collected during the COVID-19 outbreak: Lockdown 2020 (26 March - 30 May 2020); Lockdown 2021 (6 January - 7 March 2021); Roadmap 2021 (8 March - 17 May 2021); Reopen 2022 (27 January - 2 March 2022), as well as data from the classic POLYMOD study~\cite{mossong2008social}.

Our method, illustrated in Fig.~\ref{fig: Model Explanation} takes individual-level contact data from surveys together with a categorical classification of the respondent and contact (here taken to be 9 distinct age groups, but gender, occupation or sexual identity would be equally feasible), resamples the data (using machine learning to avoid distributional assumptions) to generate a larger synthetic population, and finally connects individuals to form a network. This network (and associated epidemic simulations) is then compared to: the traditional stochastic block model~\cite{holland1983stochastic} which preserves age-structure but not heterogeneity; a simpler version of our methodology which preserves heterogeneity but ignores age-structure; and a classical homogeneous model which ignores both.

We compare our realised network to the underlying survey data using a generalisation of the Wasserstein distance measure, known as the  Earth Mover's Distance~\cite{rubner1998metric}, demonstrating that our method can construct networks that are closer to the data than existing methods. We then compare epidemic simulations run on each model formulation (and for different survey data), and consider the relationships between early dynamics and the final size of an outbreak. 
Our results highlight the profound impact of heterogeneity in shaping the relationships between early and later dynamics and the optimal distribution of controls. These impacts persist even when transmission rates are scaled, accounting for the shorter duration of the most abundant contacts. We therefore conclude that existing age-structured models commonly ignore much of the information that contact surveys provide, potentially leading to erroneous epidemiological projections.

\section*{Results}
\subsection*{Network Construction}\label{sec: Network Creation}
\begin{figure}
    \centering
    \includegraphics[width=\linewidth]{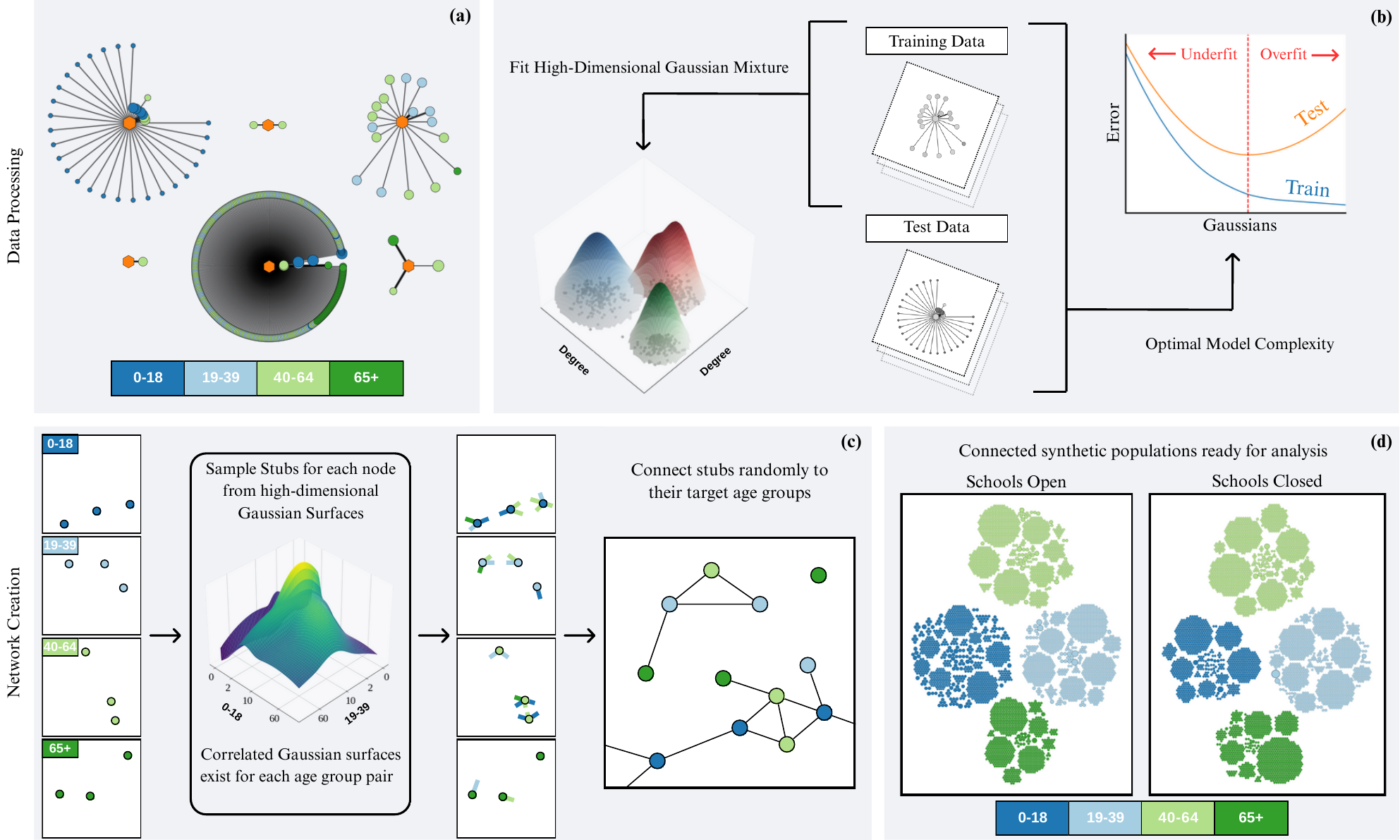}
    \caption{{\bf Steps in the network construction}: (a) Data obtained from egocentric snapshots, together with participant and contact characteristics. (b) Fit high dimensional gaussian mixtures, where each dimension represents the number of contacts with an age group for a certain duration. Find the optimal number of Gaussians which minimise the error for the predefined test set. (c) Create an unconnected population structure matching census data, sample stubs for each individual with a desired age group contact and duration. Randomly connect stubs using a stratified configuration approach. (d) Use the synthetic networks for outbreak simulation.}
    \label{fig: Model Explanation}
\end{figure}
\begin{figure}
    \centering
    \includegraphics[width=\linewidth]{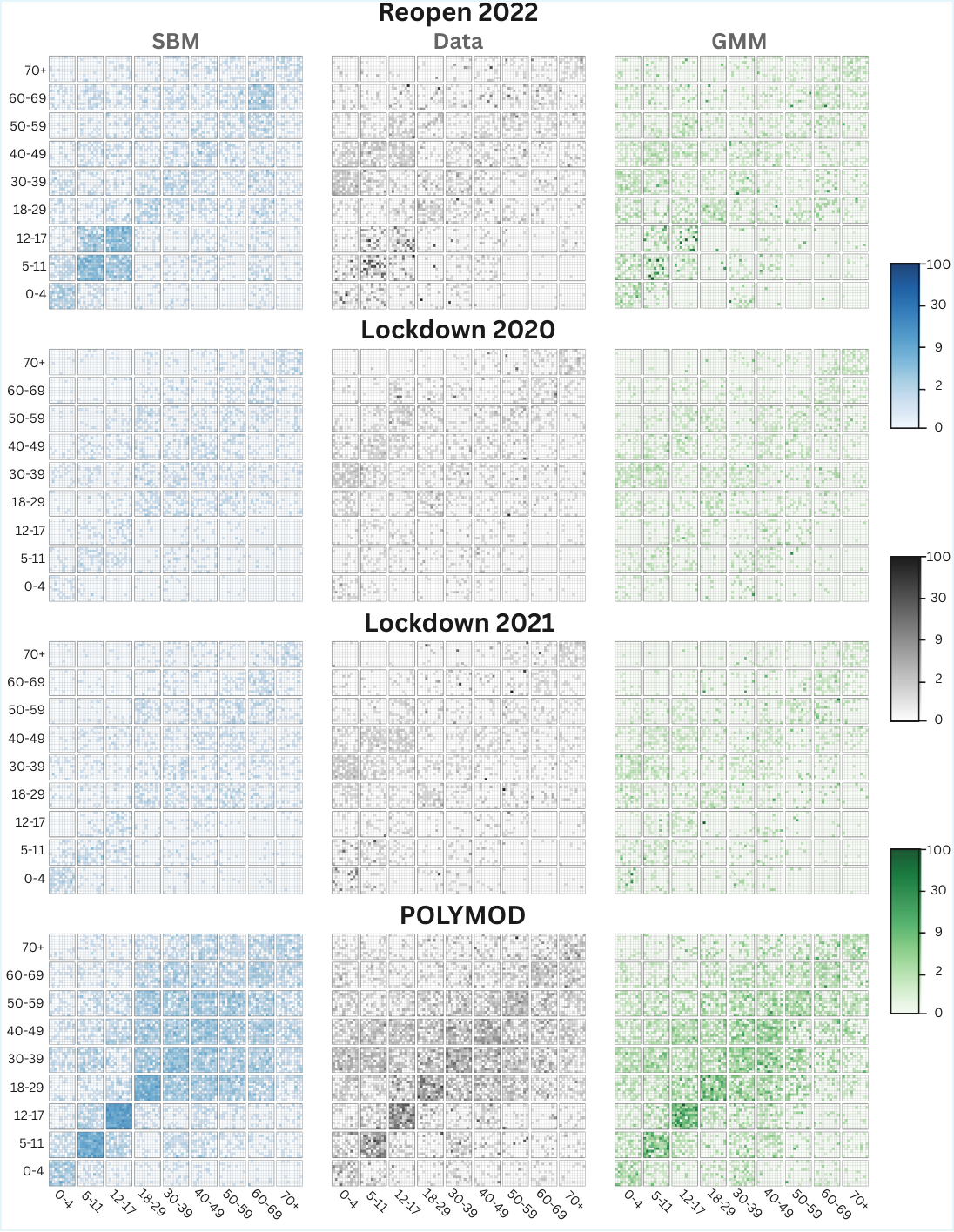}
    \caption{Contact matrices representing the mixing between age-groups and highlighting the heterogeneities in the data (grey), the stochastic block model (blue) and the GMM model (green). For the mixing between each pair of age-groups, we sample 100 ego-networks (associated with a respondent of the correct age) and calculate the number of contacts to individuals in the other age-groups. The results are then plotted as a $10\times10$ subgrid to highlight the variability - points are colour-coded on a logarithmic scale (from 0 to 100) due to the extreme heterogeneities that are present.}
    \label{fig: Contact Matrices}
\end{figure}

We define our population as a network of $N$ nodes connected by edges (throughout we take $N=100,000$), where their structure is determined by both the creation method and the underlying data set. Each node is assigned a demographic classification - here taken as one of nine age groups (0-4, 5-11, 12-17, 18-29, 30-39, 40-49, 50-59, 60-69 or 70+) years. Each edge is also associated with an interaction duration, which follows the data's classification syntax (0-5mins, 5-15mins, 15-60 mins, 1-4 hr and 4+ hr). We extrapolate our network from the respondent-level survey data through a four-step methodology (Fig.~\ref{fig: Model Explanation}). (a) We extract ego-networks (networks surrounding each respondent) from the survey data, retaining age and duration characteristics of each contact. (b) For each respondent age group we fit a finite Gaussian Mixture Model (GMM) to characterize the joint distribution of connection durations and contact age groups. For each respondent age group $a$, the GMM captures the probability of the number of connections with age group $b$ of duration length $t$ (in our case this is $9\times5=45$ dimensional).
Fitting a finite GMM requires splitting our population into training and test sets. We initially set the number of Gaussian distributions in our mixture $n_g$ equal to $1$, and perform expectation-maximization against our training data set. The Bayesian Information Criterion (BIC) provides a metric to quantify the fit  our GMM's prediction against the test data. 100 BIC scores of re-sampled test and training data gives a goodness of fit measure when using $n_g$ Gaussians. $n_g$ is then increased by 1 and the process is repeated until the average BIC begins to increase due to overfitting. The resulting values of $n_g$ can be found in Table.\ref{Tab:OptimalNumber} (c) We create a set of $N$ nodes, whose ages are chosen to match the UK census data~\cite{office2021overview}. Each node has its degree distribution sampled from the appropriate GMM, creating an ego-network of unconnected links (stubs), which have a target demographic and duration.
Sampling biases and finite sample sizes lead to inconsistencies in stub counts between age groups, such that the number of connections from age $a$ to age $b$ in the population is not the same as connections from $b$ to $a$. To overcome this error, we rescale (and stochastically round) to regain symmetry at the associated mean number. 
(d) A stratified configuration approach~\cite{molloy1995critical,guillaume2006bipartite} randomly connects stubs with compatible targets - those with the same duration and appropriate age groups. This generates our GMM network.
As a homogeneous comparator the stochastic block model (SBM)~\cite{holland1983stochastic} (with the ``communities'' in this approach defined by age classes) is used, capturing the between-age mixing without a heterogeneous degree distribution.
Fig.~\ref{fig: Contact Matrices} shows examples of the underlying age-dependent mixing matrices for the SBM, the raw data and our GMM approach, derived from the four different data sets. The heterogeneity demonstrated by the $10\times10$ sample within each age-age pair is clearly higher for the GMM model and the real data.\\

\noindent {\bf Error Metrics.} While Fig.~\ref{fig: Contact Matrices} provides qualitative evidence of the agreement between the reconstructed networks and the survey data, it is also important to quantify how closely the generated networks reproduce the underlying contact patterns. To achieve this, we can define an error between the ego-network of an individual in the data ($d_i$) and an individual in the model ($m_j$) using an Earth Mover's Distance (EMD) metric~\cite{rubner1998metric}. EMD utilizes optimal transport formulation to quantify the distance between two ego-networks, accounting for differences in the number of contacts and the ages of those contacts (see Methods for details). Intuitively, the metric quantifies the minimal number of modifications—additions, removals, or shifts in age categories—required to transform one ego-network into the other. To compute the overall network error, we select a sample of individuals whose ages and sample size agree with the data, and find the one-to-one match between individuals in the data and individuals in the model that minimises the average EMD error.

Taking 100 independent realisations of each network construction process, we can compute the average EMD per individual across the four data sets (Table \ref{Tab:EMDerror}). The ability of the GMM model to capture degree heterogeneity means that it performs substantially better than the SBM for all data sets. For the GMM networks and CoMix data, the average EMD is less than one - suggesting that the data can be perfectly reconstructed with less than one `change' (addition, subtraction, or shift of age or duration) per contact. POLYMOD data is associated with higher errors, which we partially attribute to the truncation of high numbers of contacts. The GMM model most closely captures the survey data compared to other simpler models (e.g. SBM), and including the heterogeneity of age-structure improves the fit to the overall degree distribution (the error ignoring age-structure is larger)

To test how much of this error is due to the sample size, we also perform the fitting and error estimation using a GMM network as synthetic data (which we term self-fitting). Although the error is lower
for this synthetic data, it is at least half (Lockdown 2021) to three-quarters (Reopen) of the error when GMM is fit to the data - suggesting that the majority of the EMD error is due to sampling variability rather than the inability of the GMM model to capture the degree distributions.

\begin{table} 
\begin{tabular}{ |l||c|c|c|c|}
\hline
 & \multicolumn{4}{c|}{EMD Errors for different data sets and methods} \\ 
 \hhline{~----}
Method & Reopen & Lockdown 2020 & Lockdown 2021 & POLYMOD\\
\hline
 SBM & 2.97 (2.94, 3.00) & 1.14 (1.12, 1.17) & 1.70 (1.68, 1.73) &  3.04 (3.02, 3.06)\\
 SBM no age & 4.82 (4.81, 4.84) & 1.98 (1.96, 2.01) & 3.71 (3.69, 3.73)  & 4.17 (4.12, 4.23) \\
 GMM & {\bf 0.91(0.86, 0.96)} & {\bf 0.77(0.72, 0.82)} & {\bf 0.96(0.90, 1.02)} &  {\bf 1.87(1.84, 1.90)}\\
 GMM no age & 7.94(7.91,7.97) & 7.10(7.07, 7.13)& 8.69(8.66, 8.73)& 22.2(22.05,22.37)\\
 \hline
 SBM self-fitting & 0.69 (0.67,0.71) & 0.55 (0.52,0.59) & 0.49 (0.47, 0.52) & 1.02 (1.01, 1.03) \\
 GMM self-fitting & 0.71 (0.62, 0.81) & 0.52 (0.48, 0.56) & 0.51 (0.45, 0.57) & 1.38 (1.35, 1.42) \\
 \hline
\end{tabular}
\caption{Average Earth Mover's Distance per individual comparing the model and the survey data for the four data sets and variations on the two methodologies (SBM and GMM). Reported EMD errors are the mean of 100 realisations and re-samplings of the model with the same size as the survey data. Values in brackets give the 95\% credible interval. Self-fitting uses a sample of 10,000 ego-networks from a network reconstruction, and uses this data to repeat the process above to calculate an error rate in reconstruction from a known distribution.}
\label{Tab:EMDerror}
\end{table}

\subsection*{Outbreak Simulation}\label{sec: Simulation}
With information on which models best represent our network data, we focus on how this increased realism affects epidemic simulation on the networks. To capture COVID-like behaviour, outbreaks of an SEIR-type model are simulated using the Gillespie stochastic simulation algorithm, with 3 $E$-compartments to create Erlang distributed exposed periods. We define parameters $\tau$, $1/\sigma$ and $1/\gamma$ to capture transmission rate across a contact, mean exposed period and mean infectious period. Throughout, we set $1/\sigma=3$ days and $1/\gamma=4$ days, and vary $\tau$ to achieve different early dynamics. In particular, we directly measure $R_0$ as the number of secondary cases from individuals infected in the second generation; to better capture for the role of heterogeneity in transmission, the primary infection is chosen proportional to their strength (number and duration) of contacts.

At any time $t$, the force of infection on an individual $i$ is defined as:
\begin{equation*}
    \lambda_i(t)=\tau \sum_{j\in I(t)}D_{ij}
\end{equation*}
where the sum is over all individuals that are currently infectious, and $D_{ij}$ is the duration of contact between individuals $i$ and $j$, or zero if there is no connection. When simulating, we test two methodologies one where transmission is proportional to the measured duration of infection and one in which all contacts transmit equally (equivalent to setting non-zero $D_{ij}$ to be one). While we note that physicality, proximity and setting are all likely to influence transmission risk, we use the duration of contact as a parsimonious measure. \\

\begin{figure}
    \centering
    \includegraphics[width=\linewidth]{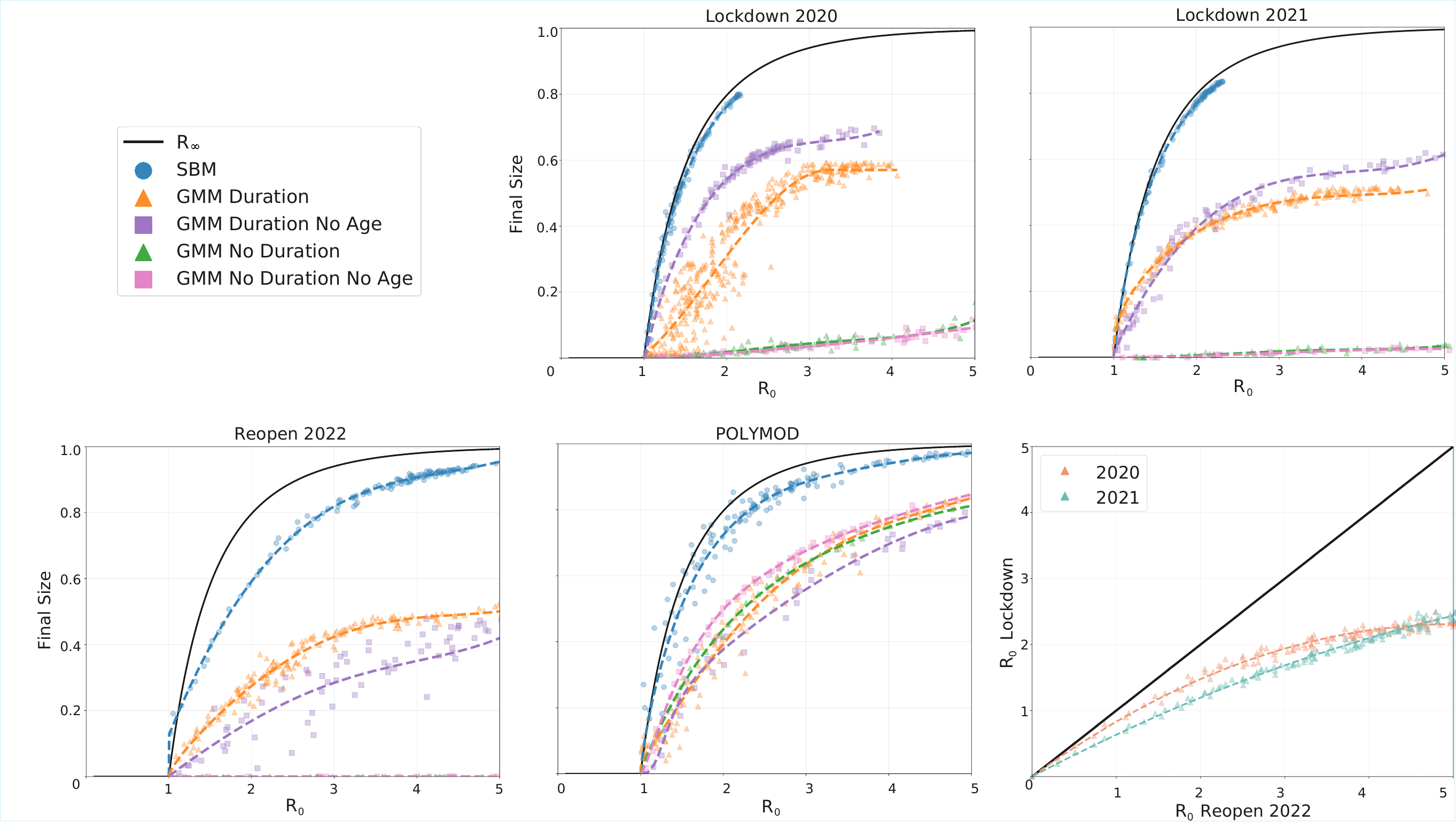}
    \caption{{\bf Relationships between $R_0$ and final size} for different survey data and different methodologies. The first four panels show final size for each data set against $R_0$ using the SBM and the GMM network with and without duration scaling. The last panel compares the $R_0$ values from CoMix data sets for the GMM model with duration scaling;  as expected lockdown networks generate lower $R_0$ values for the same transmission rate $\tau$.}
    \label{fig: final sizes for each data}
\end{figure}

\noindent {\bf Predicting outbreaks size from early dynamics.} In Fig.~\ref{fig: final sizes for each data}, the final size for a given range of $R_0$ values are compared between surveys (panels) and network construction methods (points). We note that for some constructions, it is difficult to achieve abitrarily high $R_0$, as this is bounded by the degree, as our $R_0$ is bounded by the average degree of the second generation of the outbreak minus 1. In particular, we compare the standard theoretical final epidemic size ($R_{\infty}$, black line) ~\cite{kermack1927contribution}, with results from SBM and GMM networks. For the GMM networks we further consider the impact of contact duration and age structure (given the structure of the SBM networks age and duration have minimal effect). The SBM networks (blue circles), which capture age structure but not heterogeneity are slightly reduced from the theoretical ideal, most notably outside lockdowns when the role of school-aged children may be larger.

For the more recent CoMix surveys, which more accurately capture the degree heterogeneity, ignoring the duration of contacts (green and pink) leads to vastly suppressed epidemic sizes for a range of $R_0$ values. The absence of this effect for the POLYMOD data suggests it is the action of superspreaders in the population greatly increasing $R_0$ and the early growth of outbreaks. 

When duration is included, the final size (for a given $R_0$ is higher) but still below the SBM network. The role of age-structure is complex, during the 2020 lockdown including age structure (orange) reduces the final epidemic size, but for Reopen 2022 the trend is reversed. We suggest this may be attributable to the role of school aged children. For the Reopen network, large $R_0$ values can be readily achieved but these are only associated with outbreak sizes that would reach around half the population.  

For the POLYMOD data, where very high numbers of contacts cannot be recorded, the different simulations are far closer to each other, and even ignoring duration does not substantially impact the pattern. Unexpectedly, the GMM networks display larger outbreak sizes during COVID lockdowns than when the controls are relaxed. This counterintuitive result arises because we examine the relationship between $R_0$ and epidemic size, not transmission rate ($\tau$). Lockdown controls suppressed high numbers of contacts which has a more substantial impact on $R_0$ than final size. For a given $\tau$, lockdowns lead to lower $R_0$ values (Fig.~\ref{fig: final sizes for each data}, bottom right panel) and consequently lower epidemic sizes. Interestingly, the reduction in $R_0$ due to lockdown has a different form in 2020 to 2021, which we attribute to difference in precautionary behaviour due the greater familiarity with lockdown rules by 2021.\\

\noindent {\bf The dispersion factor $k$ and secondary case distribution.} The heterogeneous transmission of infection is often characterised by the dispersion factor $k$, derived from fitting the reported numbers of secondary cases per primary case early in the epidemic to a negative binomial distribution \cite{lloyd2005superspreading,endo2020estimating}. $k$ relates to the variance of the negative binomial distribution: ${\text{Var}(NB)=R_0+R_0^2/k}$. As such, low values of $k$ are associated with with high heterogeneity and increased likelihood of superspreading events. Estimates for COVID-19 generally lie in the range 0.1-0.7 \cite{adam2020clustering, wang2021superspreading}, although these values often come from fitting to relatively sparse and possibly incomplete data. 

Estimates of $k$ derived from the early phase of our modelled outbreaks (with $R_0=1.5$) can be found in Table \ref{Tab:kdispersionfactor}. SBM networks consistently produce dispersion parameters larger than those observed for COVID-19, indicating unrealistically homogeneous transmission. In contrast, GMM networks that ignore contact duration consistently overestimate heterogeneity, producing values of $k$ that are too small for all data sets except Lockdown 2020 and POLYMOD. POLYMOD's limit of below 50 on the number of contacts a participant reports reduces the strength of high degree nodes and may explain this disparity in dynamics. When duration is included in the GMM networks, $k$ lies inside the COVID-19 observation intervals in three out of our four data sets. \\

\begin{table}
\begin{tabular}{ |l||c|c|c|c|  }
 \hline
  & \multicolumn{4}{c|}{Dispersion factor for different methods and data sets} \\
 \hhline{~----}
  & \multicolumn{2}{c|}{SBM} & \multicolumn{2}{c|}{GMM}\\
  \hhline{~----}
 Data set& No Duration & Duration & No Duration & Duration\\
 \hline
 Reopen 2022 & 1.03 (0.98,1.10) & 2.35 (2.18,2.56) & 0.07 (0.06,0.08) & $\boldsymbol{0.55 \ (0.48,0.64)}$\\
 Lockdown 2020 & 4.09 (3.65,4.62) & 2.86 (2.66,3.15) & $\boldsymbol{0.11 \ (0.11,0.13)}$& $\boldsymbol{0.66 \ (0.53,0.87)}$\\
 Lockdown 2021 & 3.81 (3.41, 4.17)& 2.58 (2.44,2.78) & 0.08 (0.08,0.09)& 1.10 (0.93,1.24)\\
 POLYMOD & 1.13 (1.08,1.18) & 1.04 (0.99,1.10) & $\boldsymbol{0.59 \ (0.55,0.62)}$& $\boldsymbol{0.56 \ (0.53,0.59)}$\\
 \hline
\end{tabular}
\label{Tab:kdispersionfactor}
\caption{{\bf k-dispersion factor of secondary case distributions} calculated from the second generation of outbreaks using an $R_0\approx1.5$. Values of $k \in [0.1,0.7]$, follow the estimated dispersion factor of COVID-19 and are in bold \cite{adam2020clustering, wang2021superspreading}. Confidence intervals are drawn from bootstrapping 100 samples from the set of secondary cases.}
\end{table}

\begin{figure}[htbp]
    \centering
    \centering
    \includegraphics[width=\linewidth]{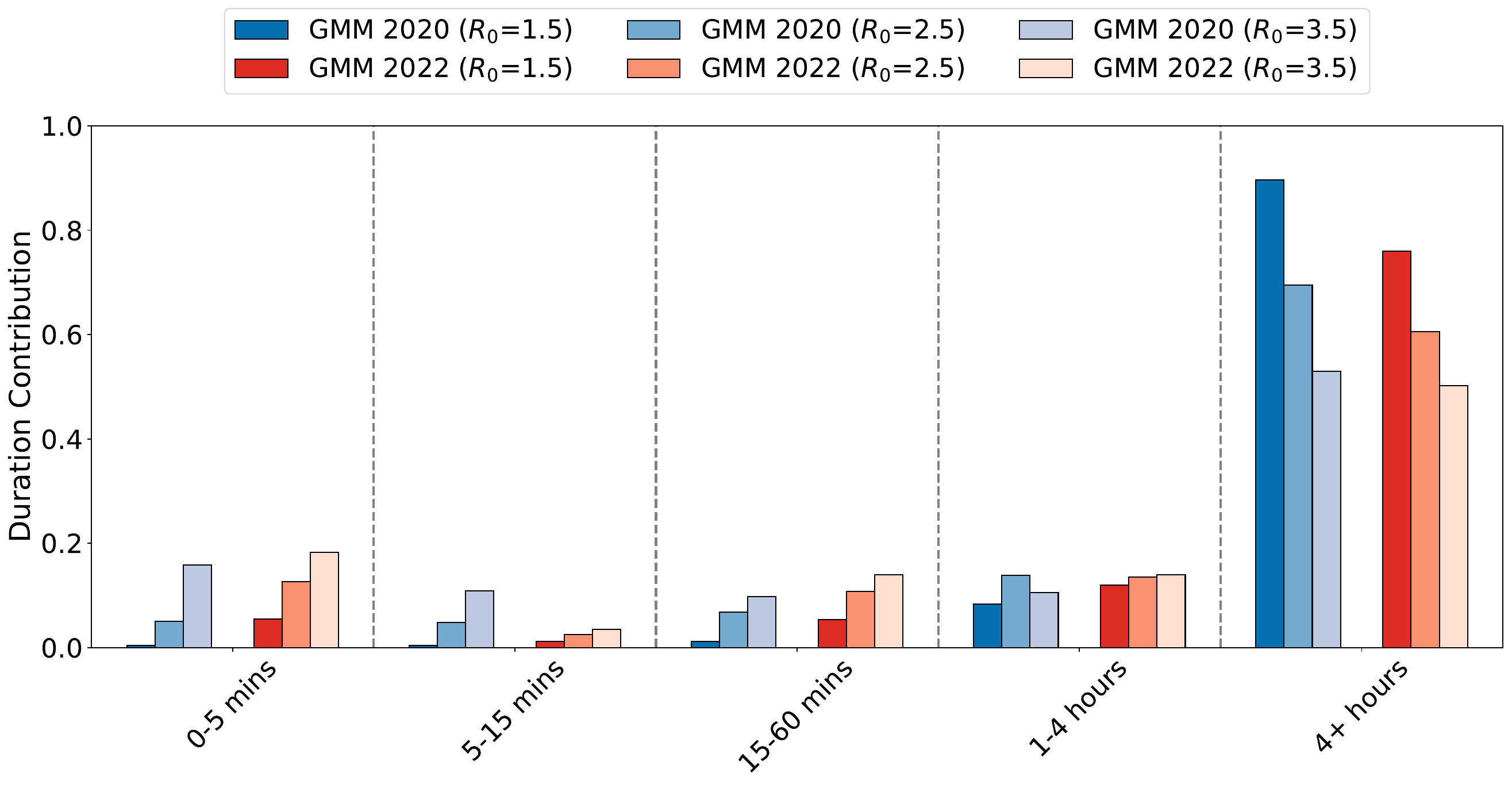}
    \includegraphics[width=\linewidth]{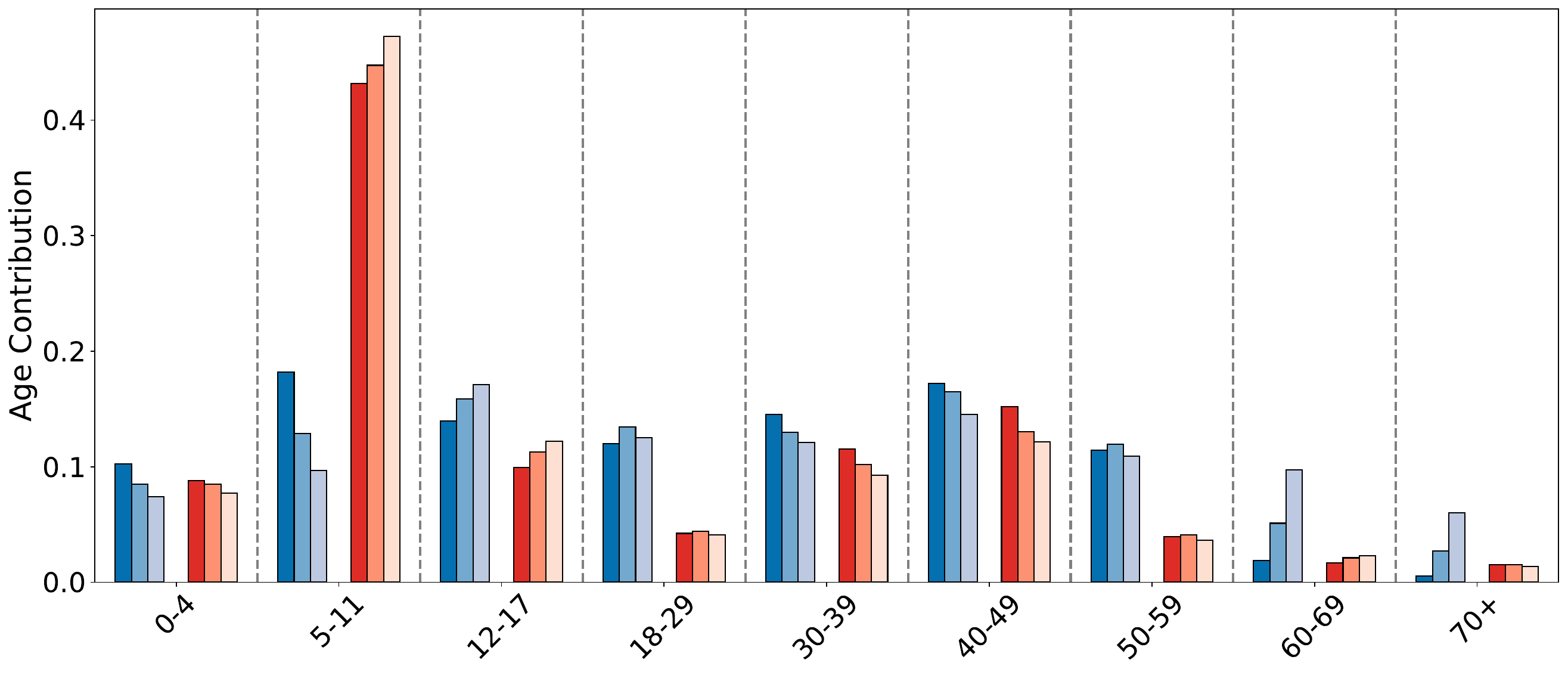}
    \caption{Contributions of different age groups and link durations to outbreaks with $R_0=1.5,\ 2.5,\ 3.5$ in each data set. Duration contributions taken as the proportion of cases from generation two caused by each link type. Age-structured contributions are calculated by the normalised leading eigenvector of age-matrix of cases from generation two.}
    \label{fig:sbm_dur_contrib}
\end{figure}

\noindent {\bf Targeting controls and early structure.} The structure embedded within the early dynamics provides important information for public health controls. In particular, for the duration-based GMM network we consider the contribution of different age groups and different contact durations to $R_0$ (Fig.\ref{fig:sbm_dur_contrib}); as such, this informs which contacts could be targeted for controls. In general, long duration contacts (4 hours or more) play the dominant role. However, for the CoMix surveys which more realistically capture high-degree nodes, very short (under 5 minute) shorter duration contacts become increasingly important as $R_0$ (and hence the transmission rate $\tau$) increases. This supports the premise that contact tracing should be targeted to long-duration contacts \cite{Kendal_2024}, but also suggests that random short-duration and largely untraceable contacts could maintain transmission if $R_0$ was sufficiently high \cite{Keeling_Holthingsworth_Read_2020}. 

The age-structured patterns are more complex with multiple subtle trends emerging. We consistently find that older age-groups contribute the least to $R_0$, while school age children (predominantly 5-11 year olds) and 30-49 year olds contribute the most. This pattern is damped during the first lockdown in 2020, but becomes stronger in 2021; once the schools are reopen in 2022 then 5-11 year olds contribute over 40\% to the early growth of infection. This suggests that closing primary schools may have a substantial impact on future outbreaks, although these are associated with considerable social and educational disruption.
We note that for POLYMOD it is 12-17 year olds that dominate (see supplementary material), although it is unclear if this is a long-term or temporary change driven by the pandemic. 
Comparisons with SBM networks (see supplementary material) indicate a reduced contribution from younger age groups, suggesting that the greatest contact heterogeneity occurs within these populations.

\section*{Discussion}\label{sec13}

Here, we present a general methodology for the construction of age- and duration-structured contact networks from ego-centric network data (Fig.~\ref{fig: Model Explanation}). This methodology extrapolates from the sampled respondents to generate a population-scale network while preserving the observed patterns of age-dependent mixing and contact durations. This process allows us to approximate routinely reported power-law distribution of contacts in social systems~\cite{adamic2001search,csanyi2004structure}, or alternate distributions \cite{broido2019scale} without user input in the fitting process. Throughout, we compare our network with that derived from the Stochastic Block Model (SBM)~\cite{holland1983stochastic} which ignores heterogeneity -- leading to Poisson degree distributions -- while retaining age-structure.

We have considered four exemplar data sets that provide ego-centric network information from the UK: the ground-breaking POLYMOD study~\cite{mossong2008social}; and three snapshots from the CoMix survey~\cite{gimma2022changes} taken during the COVID-19 pandemic, with qualititatively different social restrictions. Our error metric -- comparing ego-networks from the data with those from our synthetic network -- highlights the need for a heterogeneous approach and the power of using the flexible Gaussian Mixture Model (GMM) to capture the full distribution of contacts recorded in the CoMix data (Table \ref{Tab:EMDerror}). 
It should be noted that while these contact surveys are state-of-the-art in terms of quantifying human contacts, they are not necessarily a perfect reflection of epidemiologically important contacts. For example, they may suffer from issues with recall-bias and estimating the age of contacts~\cite{mccormick2007adjusting,voelkle2012let}. Also, such surveys commonly record data on face-to-face conversational contacts, and while this is a good proxy for epidemic risk for infections spread through close contact, it misses long-term co-location which could play a role in long range air-borne transmission.

We simulated epidemic outbreaks on our networks based on their realised early dynamics, as captured by $R_0$ (although basing our analysis on early growth rate generates similar results),  and show that epidemic outcomes are heavily influenced by network heterogeneity. When all contacts are treated equally, the behaviour of the more accurate GMM network is dominated by rare highly-connected superspreaders, who increase $R_0$ without leading to substantially larger outbreaks. In reality, the risk of transmission is likely to be positively correlated with the duration of a contact, and the CoMix data clearly shows that average duration declines with the number of contacts. By assuming that transmission risk is proportional to duration, our model is able to capture the heterogeneous secondary case distribution that has been observed for COVID-19 and other infectious diseases, often characterised using a negative binomial~\cite{danon2012social, endo2020estimating}. A wealth of other factors are likely to influence the risk of transmission across a contact, including the intimacy of contact and the setting in which it occurs~\cite{ferretti2024digital}, but such information is difficult to gather and hard to robustly quantify.

The relationships between $R_0$ and epidemic size for the different models demonstrate how age-structure and degree heterogeneity both shape an outbreak. When considering the final epidemic size, our models suggest that age-structure plays a limited role (once simulations are matched to the same initial $R_0$). However, for many important infectious diseases (e.g. influenza or COVID-19) a population average is not a useful measure, especially when disease severity is strongly age-dependent~\cite{davies2020age}, or when age-related interventions, such as school closures, require careful evaluation~\cite{andersen2020early}. Such age-dependent factors have been ignored here for simplicity, but could be readily incorporated into any public health focused model.
Degree heterogeneity is predicted to have a striking effect, even when it is moderated by the average duration -- which declines with increasing degree. The diversity between our results highlights how higher order network structures, not captured by $R_0$ (nor other early measures of epidemic growth) can profoundly impact the course of an outbreak. 

From a survey design perspective, our results demonstrate the need for robust survey designs that capture the full heterogeneous nature of social contacts~\cite{danon2011networks}. In particular, there is a clear difference in contact distributions reported by the CoMix~\cite{gimma2022changes} and POLYMOD~\cite{mossong2008social} surveys. It remains an open and challenging question how much of this difference is due to the survey methodology (with a limit on the number of contacts reported in POLYMOD), or how much is due to temporary or permanent changes in social mixing post-COVID restrictions. More data is needed to characterise the epidemiological role of highly-connected individuals and to understand if the well-used POLYMOD survey still holds in a post-COVID setting.

As with any modelling approach, we have made approximations to gain a tractable approach. Potentially the five main issues are: (i) neglecting the context and intimacy of the connections, which during the COVID-19 pandemic could include one or both individuals wearing masks. (ii)  The assumption that the network still holds when individuals are ill, whereas highly symptomatic individuals are likely to be forced to reduce their contacts. Hence, our assumptions are most robust when the infection has a pronounced prodromal infectious period. (iii) Our network construction is unlikely to lead to clusters (triangle-forming contacts), yet we intuitively expect many clustered connections in household, work and leisure settings~\cite{danon2012social}. Including clustering in a data-driven way is extremely difficult, as it requires survey participants to estimate information about the contacts of their contacts \cite{danon2012social}. However, clustering is epidemiologically important as it has a dampening effect on epidemic growth~\cite{house2011epidemic}. (iv) Finally, we have assumed a constant network structure based on a daily snapshot of individual behaviour; while many contacts (such as household contacts) are repeated daily, others may have a weekly pattern or be irregular. Incorporating this structure into our networks requires the addition of an extra dimension (frequency of contact), but even this would neglect the potential daily variability which would require repeated sampling to quantify~\cite{beraud2015french}. (v) Finally, we have used COVID-like parameters as a motivating example, but have not included the rich epidemiology associated with this infection -- such as age-dependent severity and infectivity -- hence our results are representative of generic epidemiological dynamics rather than aiming to provide robust public health projections for COVID-19.

We have generated a robust approach for generating and validating accurate structure networks, which can be used for epidemic simulation. Our findings have important implications for both public health modelling and future survey design; highlighting the importance of capturing heterogeneity, the need to understand clustering and the regularity of behaviour, and the urgent need for new surveys to bridge the gap between POLYMOD and post-COVID mixing patterns. Our analysis was focused on UK data sets comprising differing periods of social restrictions, but there are a large and growing number of contact survey data sets available from different countries, many of which use the same syntax. The methodology described here could be applied to any of these data sets, allowing for out-of-the-box application to any population of choice and an exploration of epidemiological differences. 

\newpage

\section*{Methods}
\subsubsection*{Data Preprocessing}
Our data are drawn from anonymised, publicly available survey datasets, curated and stored in a consistent format on \href{https://socialcontactdata.org}{socialcontactdata.org}. In this study, we use only participant age, contact age, and contact duration, deliberately excluding other potential covariates such as household size, gender, and contact frequency. Participant age is assumed to be known exactly. In contrast, contact age in the CoMix study \cite{gimma2022changes} is reported as an interval, with widths that vary across responses. When a reported age range lies entirely within a single predefined age-demographic bucket and spans fewer than 15 years, we sample uniformly from that range. Broader age ranges, such as 0–17 or 18–64, span multiple demographic groups. For these, after sampling narrower ranges, we mitigate bias by assigning contacts to age buckets with probabilities proportional to the age-specific mixing patterns observed in the data—that is, conditional on the distribution of contacts an average individual of a given age has with each age group. When contact age is missing, we apply the same procedure used for broad age ranges, allowing all age buckets. In contrast, POLYMOD \cite{mossong2008social} reports exact contact ages, eliminating the need for age-distribution assumptions. Finally, contacts with missing duration information are classified as short-duration contacts (0–5 minutes).

\subsubsection*{Network Creation}
Each respondent is represented as an ego network in which all nodes are annotated with age and each edge is associated with a contact duration. We encode each ego network $i$ as a 45-dimensional vector $\boldsymbol{d_i}$, capturing the ego’s contact frequency distribution. The 45 dimensions correspond to all pairwise combinations of 9 age brackets and 5 contact-duration categories. Prior to model fitting, we apply a logarithmic transformation to the ego-network vectors in order to mitigate the influence of heavy-tailed contact distributions. Specifically, each component of $d_i$ is transformed as
\begin{equation*}
    \boldsymbol{d'_i}=\log(\boldsymbol{d_i}+\boldsymbol{1}),
\end{equation*}
where the logarithm is applied elementwise and $\boldsymbol{1}$ denotes a vector of ones. The additive offset ensures that zero-valued components are well-defined under the transformation. This log-transform compresses large values while preserving relative differences among smaller counts, yielding a distribution that is better approximated by a finite mixture of Gaussian components using the Earth Mover's Distance error metric.

Let $A$ denote our predefined set of age groups and $\mathcal{D}_a=\{\boldsymbol{d'_i}\}_{i\in a}$ denote the resulting set of observed vectors from egos in $a\in A$. We model $\mathcal{D}_a$ using a finite Gaussian mixture model (GMM) and estimate its parameters via the Expectation–Maximization (EM) algorithm by maximising the log-likelihood
\begin{equation*}
    \ell(\theta) = \sum_{i=1}^{N_a}\log\left(\sum_{j=1}^{n_g}\phi_j\mathcal{N}\left(\boldsymbol{d'_i}|\boldsymbol{\mu_j},\boldsymbol{\Sigma_j}\right)\right),
\end{equation*}
where $N_a$ is the number of observations and $n_g$ is the number of Gaussian components. The mixture weights $\phi_j$ satisfy $\sum_{j=1}^K\phi_j=1$, and $\boldsymbol{\mu_j}$ and $\boldsymbol{\Sigma_j}$ denote the mean vector and covariance matrix of component $j$, respectively.

To select an appropriate number of mixture components, we randomly partition $\mathcal{D}_a$ into training (80\%) and test (20\%) sets. For a fixed value of $n_g$, the GMM is fitted to the training data using EM, and the Bayesian Information Criterion (BIC) is evaluated on the test set. This procedure is repeated over 100 independent random train–test splits. The mean BIC across all partitions provides a measure of model suitability for a given $n_g$. We select the value of $n_g$ that minimises the mean BIC and use the resulting GMM, $\mathcal{M}_a$, to represent the corresponding data subset $\mathcal{D}_a$. The set $\{\mathcal{M}_a\}_{a\in A}$ defines the collection of age-specific contact structure models used to build networks for subsequent simulations. 

\begin{algorithm*}[!ht]
\caption{Generate GMM Network}\label{algo: My Model}
\begin{algorithmic}[1]
\Require $n \geq 0$
\Ensure $A \Leftarrow$ Age-Duration-Specific-Network$(n, \mathcal{D})$
\State params $\Leftarrow$ get-GMM-fits($\mathcal{D}$) \Comment{Store GMM fits for each age-group}
\State $A \Leftarrow \text{matrix(n,n)}$ \Comment{Define adjacency matrix of zeros}
\While{$C_a \in \boldsymbol{C}$}
    \While{$C_b \in \boldsymbol{C}$} \Comment{Loop through all age group combinations}
        \While{$i \in C_a$}
            \State $k_a[i] \Leftarrow$ sample-dist(params$,a,b$) \\ \Comment{Generate stubs for $i$ given params of block $a\to b$}
        \EndWhile
        \While{$j \in C_b$}
            \State $k_b[j] \Leftarrow$ sample-dist(params$,b,a$) \\ \Comment{Generate stubs for $j$ given params of block $b\to a$}
        \EndWhile
        \State $\boldsymbol{k_a},\boldsymbol{k_b} \Leftarrow$  
        \While{$\boldsymbol{k_a} \neq [\ ]$ and $\boldsymbol{k_b} \neq [\ ]$} \Comment{While degrees vectors are not empty}
            \State $\boldsymbol{k_a}, \boldsymbol{k_b} \Leftarrow$ remove-zeros($\boldsymbol{k_a},\boldsymbol{k_b}$) \\ \Comment{Remove nodes with 0 stubs from choice}
            \State $\boldsymbol{k_a}, \boldsymbol{k_b} \Leftarrow$ remove-complete($\boldsymbol{k_a}, \boldsymbol{k_b}$) \\ \Comment{Remove nodes connected to all remaining targets} 
            \State $i \Leftarrow $ random-index($k_a$) \Comment{Random index of $a$ with non-zero degree}
            \State $j \Leftarrow$ random-index($k_b[A_{ij} = 0]$) \\ \Comment{Random index of $b$ without connection to $i$}
            \State $A_{ij} \Leftarrow 1$ \Comment{Assign link}
            \State $k_a[i] \Leftarrow k_a[i] - 1$ \Comment{Reduce the number of stubs leaving $i$}
            \State $k_b[j] \Leftarrow k_b[j] - 1$ \Comment{Reduce the number of stubs leaving $j$}
        \EndWhile
    \EndWhile
\EndWhile
\State $\Rightarrow A$ \Comment{Return network adjacency matrix}
\end{algorithmic}
\end{algorithm*}

We represent the population as a set of $n$ nodes, each with an age attribute. Each individual $i$ is assigned a vector of yet unconnected half (stubs) 
\begin{equation*}
    \boldsymbol{z_i}\sim\sum^{n_g(a)}_{j=1}\phi_{j}^{(a)}\mathcal{N}\left(\boldsymbol{\mu}_{\boldsymbol{j}}^{(a)},\boldsymbol{\Sigma}_{\boldsymbol{j}}^{(a)}\right),
\end{equation*} 
for $i\in a$. Each value in $\boldsymbol{z_i}$ refers to the number of contacts $i$ wants with a certain age-duration category. 

\begin{table}
\begin{tabular}{| l||c|c|c|c|}
 \hline
  & \multicolumn{4}{c|}{Optimal Number of Gaussian Components $n_g$} \\
  \hhline{~----}
  & \multicolumn{2}{c|}{Age} & \multicolumn{2}{c|}{No Age} \\
  \hhline{~----}
  Data set& Duration& No Duration& Duration &No Duration\\
 \hline
 Reopen 2022 & [1,4,3,7,3,7,4,2,3]& [1,6,7,11,11,9,9,15,6]& 77 & 27\\
 Lockdown 2020 & [1,1,2,3,3,5,1,1,1]& [4,5,7,9,8,16,15,9,12]& 71 & 20\\
 Lockdown 2021 & [2,2,2,2,3,7,4,6,1]& [5,11,13,15,16,16,19,15,14]& 69 & 25\\
 POLYMOD & [1,4,3,7,3,7,4,2,3]& [1,6,7,11,11,9,9,15,6]& 77 & 27\\
 \hline
 \hline
\end{tabular}
\label{Tab:OptimalNumber}
\caption{{\bf Optimal number of Gaussian Components $n_g$} fitted using EM algorithm. In age-structured simulations $|A|=9$, leading to a 9 dimensional vector $\boldsymbol{n_g}$ in ascending age order. Otherwise, $n_g$ is an integer valued number of components for entire data set.}
\end{table}

Network construction proceeds by extending the configuration model to incorporate both age and duration structure. Stubs are paired only when their attributes are compatible: both ends of a link must specify the same contact duration, and the target age group of each stub must match the age of its counterpart. In practice, perfect matching is not always possible due to asymmetries in reported contacts between age groups. Such inconsistencies may arise from biased population sampling—leading to overrepresentation of certain demographics—or from systematic misreporting of interactions. As a result, the realised number of links between two age groups satisfies
\begin{equation*} \left|\text{links}(a\to b)\right|\approx \min\left(\text{stubs}(a\to b), \text{stubs}(b\to a)\right),
\end{equation*}
with increasing asymmetry leading to reduced overall connectivity. 

To correct for directional asymmetries in contact counts, we rescale contacts in the direction with fewer reported links. We first define the total number of stubs from age group $a$ to age–duration category $x$ as
\begin{equation*}
    L_{a\to x} = \sum_{i\in a}z_i(x).
\end{equation*}

We then define the reciprocal quantity $L_{a\to x}^{(-1)}$, which counts the number of links in the opposite direction, from the target age group back to the source age group $a$, for the same contact duration. 

For each age-duration pairing $(a,x)$, we define the scaling factor
\begin{equation*}
    s_{ax} = \frac{L_{a\to x}+L_{a\to x}^{(-1)}}{2\min(L_{a\to x},L_{a\to x}^{(-1)})}.
\end{equation*}
The scaling factor satisfies $s_{ax}=1$ when directional stub counts are symmetric and increases monotonically with the degree of asymmetry. Stub vectors are therefore updated as
\begin{equation*}
    z'_i(x) =  
    \begin{cases}
        s_{ax}z_i(x)&\text{, if } L_{a\to x} < L_{a\to x}^{(-1)},\\
        z_i(x) &\text{, otherwise.}
    \end{cases}
\end{equation*}
Since this rescaling generally produces non-integer stub counts, we apply stochastic rounding to obtain integer-valued vectors while preserving expectations. For each component,
\begin{equation*}
    z'_i(x)=
    \begin{cases}
        \left\lfloor z'_i(x)\right\rfloor, \text{   with probability } \left\lceil z'_i(x)\right\rceil - {z'_i}(x),\\
        \left\lceil z'_i(x)\right\rceil, \text{ with probability } z'_i(x) - \left\lfloor z'_i(x)\right\rfloor.
    \end{cases}
\end{equation*}
This ensures $\mathbb{E}[z'_i(x)]$ equals the rescaled value, maintaining consistency with the configuration model while restoring integrality.

Once stubs are paired to form edges, we assign weights to links according to contact duration. For each connected pair of nodes $i$ and $j$ with duration category $\tau$, we define

\begin{equation*}
    D_{ij} = \frac{f(\tau)}{\max_{\tau'}f(\tau')},
\end{equation*}
where $f(\tau)$ denotes the representative duration associated with category $\tau$ and $\boldsymbol{D}$ is the adjacency matrix of our network. In the empirical data, $f(\tau)$ is taken as the mean of the corresponding duration interval. For the open-ended category “4+ hours”, we assume an upper bound of 12 hours, yielding
\begin{equation*}
    f(\text{4+ hours}) = \text{8 hours}.
\end{equation*}
This normalisation ensures edge weights lie in $[0,1]$ and reflect relative contact intensity.

The Stochastic Block Model (SBM) provides a simple mechanism for inducing age-based community structure in a population, but it does not directly extend to networks with duration-weighted contacts. An SBM is defined by its edge-probability matrix $\boldsymbol{P}$, where 
\begin{equation*}
    P_{ab}=\frac{C_{ab}p_b}{n},
\end{equation*}
denotes the probability that a link exists between a node in age group $a$ and a node in age group $b$. Here, $\boldsymbol{C}$ is the empirical contact matrix, with $C_{ab}$ representing the average number of contacts a respondent in age group $a$ reports with individuals in $b$; $p_b$ is the proportion of the population belonging to age group $b$; and $n$ is the total population size. This formulation preserves overall link density and the observed age-specific mixing patterns.

However, sampling and reporting biases in the data may introduce asymmetry in $\boldsymbol{C}$, which propagates to $\boldsymbol{P}$. For a node $i \in a$, the number of connections with age group $b$,
\begin{equation*}
    k_{ib} \sim \text{Bin}(np_b,P_{ab}),
\end{equation*}
follows a binomial distribution. When $n\gg1$, we have $P_{ab}\to0$ while $np_bP_{ab}=C_{ab}$ remains fixed. Under these conditions, the binomial distribution converges to a Poisson distribution,
\begin{equation*}
    k_{ib}\sim \text{Pois}(C_{ab}).
\end{equation*}
Consequently, the total degree distribution within each age group can be expressed as a sum of $|A|$ independent Poisson variables. While analytically convenient, this construction struggles to reproduce the heavy-tailed degree distributions frequently observed in empirical social networks.

Incorporating contact duration does not naturally generalise within the standard SBM framework, since each node must belong to a single community. Assigning both age and duration as community labels would imply overlapping memberships. Mixed Membership Stochastic Block Models \cite{airoldi2008mixed} address this limitation, but for simplicity we adopt a more direct extension.

Edges are first drawn according to the SBM mechanism, with
\begin{equation*}
    A_{ij} \sim \text{Bernoulli}(P_{ab}), \space\space i\in a, j \in b.
\end{equation*}
We then decompose the contact matrix into duration-specific components. Specifically, we assume
\begin{equation*}
    C_{ab} = \sum_\tau C_{ab}^{(\tau)},
\end{equation*}
where $C_{ab}^{(\tau)}$ denotes the average number of contacts from age group $a$ to $b$ with duration category $\tau$. Conditional on the existence of an edge between $i\in a$ and $j \in b$, its duration is assigned according to 
\begin{equation*}
    \mathbb{P}[A_{ij} = f(\tau)|A_{ij}\neq0] = \frac{C_{ab}^{(\tau)}}{C_{ab}}, \space\space i \in a,j\in b,
\end{equation*}
This preserves the underlying SBM age-mixing structure while incorporating duration heterogeneity in a probabilistically consistent manner.

\subsubsection*{Network Accuracy}
The quality of representative networks is often measured using overall population metrics, such as the degree distribution. In reconstruction of age-structured disease dynamics many non-linear and more granular properties of the network become important \cite{galvani2005dimensions}. This motivates the use of an error metric that quantifies similarity at the level of individual contact patterns. 

We define an {\it ego-network} as the subgraph consisting of a focal individual (the ego) and all of their direct connections. Ego-networks provide a natural unit of comparison, as they match the granularity of the survey data: each respondent can be represented as an ego-network aggregated over the reporting period. Evaluating similarity at this level allows us to assess whether synthetic networks reproduce empirically observed contact structures.

To quantify similarity between ego-networks, we employ the Earth Mover’s Distance (EMD) \cite{rubner1998metric}, a generalisation of the first-order Wasserstein distance ($\boldsymbol{W}_1$) \cite{kantorovich1960mathematical, vaserstein1969markov}. EMD is formulated as an optimal transport problem. Let $\boldsymbol{d}_i$, $\boldsymbol{k}_i \in \mathbb{R}_{\geq0}^{|A|}$ denote two discrete distributions over $|A|$ age categories, representing two ego-networks ($\boldsymbol{d}_i$ from the data and $\boldsymbol{k}_i$ from the surrogate network). Let $\boldsymbol{Q} = (q_{ab})\in \mathbb{R}_{\geq0}^{|A|\times|A|}$ be a cost matrix, where $q_{ab}$ denotes the ground distance between age categories $a$ and $b$. 

We define  a transport plan $\boldsymbol{T}=(T_{ab})\in\mathbb{R}_{\geq0}^{|A|\times|A|}$, where $T_{ab}$ represents the mass transported from age-category $a$ in $\boldsymbol{d}_i$ to age-category $b$ in $\boldsymbol{k}_i$. The Earth Mover's Distance is then given by
\begin{equation*}
    \text{EMD}(\boldsymbol{d}_i, \boldsymbol{k}_i) = \min_{\boldsymbol{T}\geq0}\sum_{a=1}^{|A|}\sum_{b=1}^{|A|}T_{ab}d_{ab},
\end{equation*}
subject to the marginal constraints
\begin{equation*}
    \sum_{b=1}^{|A|}T_{ab}\leq d_{ia}, \space \space\space \sum_{a=1}^{|A|}T_{ab}\leq k_{ib}, 
\end{equation*}
and
\begin{equation*}
    \sum_{a=1}^{|A|}\sum_{b=1}^{|A|} T_{ab} = \min\left(\sum_{a=1}^{|A|}d_{ia}, \sum_{b=1}^{|A|}k_{ib}\right).
\end{equation*}
When $\sum_ad_{ia}=\sum_bk_{ib}$, this reduces to the classical Wasserstein-1 distance. The relaxed marginal constraints allow comparison of distributions with differing total mass, which is essential when individuals report different total numbers of contacts.

In our setting, each ego-network is represented as a discrete frequency distribution over age categories, and the cost matrix $\boldsymbol{Q}$ encodes the dissimilarity between categories (for example, the distance between age groups). EMD therefore provides a principled and interpretable measure of structural discrepancy at the individual level.

In principle, duration discrepancies can also be incorporated into $\boldsymbol{Q}$ by defining a joint age–duration ground metric. However, doing so requires specifying the relative weighting between errors in reported age, reported duration, and inaccuracies arising from omitted or spurious contacts. In the absence of a data-driven basis for calibrating these relative contributions, we restrict $\boldsymbol{Q}$ to age-based dissimilarities and leave a fully integrated age–duration metric to future work.

Participant age, ethnicity and facial expressions are important contextual characteristics in the accuracy of survey-based age estimation \cite{burt1995perception, rhodes2012own, voelkle2012let}. However, in the absence of a principled quantification of these confounding effects, we define distances between consecutive age groups to be uniform in a Euclidean metric space.

We define the ground cost matrix $\boldsymbol{Q}^{(i)} \in \mathbb{R}^{|A|\times|A|}$ for respondent $i$ in the data as 
\begin{equation*}
    Q_{ab}^{(i)} = \frac{|a-b|}{||\boldsymbol{d}_{i}||},
\end{equation*}
Where $||\cdot||$ denotes the degree of individual $i$ or the total mass of the frequency distribution of this data point. Thus, $\boldsymbol{Q}^{(i)}$ takes the form 
\begin{equation*}
\boldsymbol{Q}^{(i)}=\frac{1}{||\boldsymbol{d}_i||}
 \begin{bmatrix}
0 & 1 & 2 & ... & A-1 \\
1 & 0 & 1 & \dots & A-2 \\
2 & 1 & 0 & \dots & A-3 \\
\vdots & \vdots &\vdots & \ddots & \vdots \\
A-1 & A-2 & A-3 & \dots & 0 
\end{bmatrix} .
\end{equation*}
This specification assumes a linear geometry over age categories and disregards potential correlation between contact age and survey reporting error. To reduce sensitivity to extreme values in the tails of the degree distribution, we normalise total mass. When comparing ego-network distributions $\boldsymbol{d}_i$ (data) and $\boldsymbol{k}_j$ (model), we penalise excess unmatched mass using
\begin{equation*}
    \text{penalty}(i,j) = \frac{c}{\left(\frac{||\boldsymbol{d}_i|| + ||\boldsymbol{k}_j||}{2}\right)+1},
\end{equation*}
where $c=10$. This penalty is independent of the structural arrangement of edges and depends only on aggregate degree magnitude, thereby remaining agnostic to node connectedness. 

In implementation $\boldsymbol{Q}^{(i)}$ controls the scale of transportation costs and does not alter total mass in either distribution. It only determines the cost of reallocating mass between categories. Whereas, the penalty addresses a different issue; the inequality of total mass between the two ego-network distributions, where $\boldsymbol{W}_1$ breaks down. It quantifies the cost of excess mass and is not dependent on where the discrepancy lies. 

The Earth Mover’s Distance is computed using the fast implementation of Pele and Werman \cite{pele2009fast}, which formulates optimal transport as a minimum cost flow problem on a bipartite network. 

A random sample of ego-networks is drawn from each synthetic population network such that the sample matches the empirical data in both size and age composition. We assume no uncertainty in survey participants’ reported ages, so comparisons are restricted to individuals within the same age group.

For participants indexed by $i$ and sampled model ego-networks indexed by $j$, we define the cost matrix $\boldsymbol{E}$ as
\begin{equation*}
    E_{ij} = \begin{cases}
        \text{EMD}(\boldsymbol{d}_i,\boldsymbol{k}_j)&, \text{ if } i,j \in a,\\
        \infty &, \text{ otherwise,}
    \end{cases}
\end{equation*}
where $\boldsymbol{d}_i$ denotes the empirical ego-network of participant $i$, $\boldsymbol{k}_j$ denotes the sampled model ego-network $j$, and $a$ indicates membership in the same age category. Assigning infinite cost across age groups enforces strict age-stratified matching.

The total reconstruction error is then formulated as a linear sum assignment problem, equivalently a minimum-weight perfect matching in a bipartite graph. Letting $\pi$ denote a permutation over individuals within age group $a$, we solve
\begin{equation*}
    \min_\pi\sum_{i\in a}E_{i,\pi(i)}.
\end{equation*}
This optimisation assigns each participant to exactly one model ego-network, and each model ego-network to exactly one participant, ensuring a one-to-one correspondence. Because the number of sampled ego-networks equals the number of participants within each age group, a perfect matching exists.

The problem belongs to a well-studied class of constrained linear optimisation problems and can be solved efficiently using the algorithm described by Crouse \cite{crouse2016implementing}, an implementation of the Jonker–Volgenant variant of the Hungarian method. In this context: Participants act as ``agents", sampled ego-networks act as ``tasks," and each agent and task is used at most once. 

The minimal objective value provides the total age-stratified reconstruction error. Dividing by the number of participants yields the average EMD error per individual.

To quantify the intrinsic reconstruction error induced purely by stochastic network generation and node sampling (under the assumption that the model class is correctly specified) we define a baseline error estimation procedure. This isolates variability arising from finite sampling and network realisation, rather than model misspecification.

We first fit the model to the empirical data and construct a large surrogate network of size $n$, where $n \gg n_s$ and $n_s$ denotes the size of the ego-network sample used for evealuation. The surrogate network is intended to approximate the model’s population-level distribution with negligible finite-size effects. In our implementation, we take $n=100,000$.

From this surrogate network, we draw a stratified random sample of $n_s=10,000$ nodes to form a temporary data set. The number of sampled nodes in each age group follows $\lceil n_s\boldsymbol{p}\rceil$, where $p$ is the vector of census-based age proportions and $\lceil\cdot\rceil$ is applied componentwise. Sampling is performed without replacement within each age group. The resulting collection of ego-networks is treated as synthetic ``observed” data drawn from the fitted model.

We then refit the same model to this temporary data set and generate a second surrogate network. The reconstruction procedure described previously based on age-stratified EMD and optimal bipartite matching is applied to compute the minimal transport discrepancy between the temporary data and the newly generated surrogate network. 

Because both the temporary data and the reconstructed network are generated from the same model family, the resulting EMD represents the irreducible baseline error attributable to stochastic sampling and network construction. Repeating this procedure across multiple independent surrogate networks yields an empirical distribution of this baseline error. This provides a principled reference level against which reconstruction errors obtained from the true survey data can be compared, allowing us to distinguish model inadequacy from unavoidable sampling variability.

\subsubsection*{Simulation Method}
The epidemic process is modelled as a stochastic SEIR-type system simulated using the Gillespie algorithm, representing a continuous-time Markov jump process on the network. We define the process based on the stochastic rates acting on an individual (node) $i$ within the network. 
\begin{eqnarray}
\mbox{Rate}(S \to E_1 | i = S) & = & \lambda_i(t) \\
\mbox{Rate}(E_1 \to E_2 | i = E_1) & = & 3\sigma\\
\mbox{Rate}(E_2 \to E_3 | i = E_2) & = & 3 \sigma \\
\mbox{Rate}(E_3 \to I | i = E_3) & = & 3 \sigma \\
\mbox{Rate}(I \to R | i = I) & = & \gamma
\end{eqnarray}

Here, the exposed period is modelled using three sequential compartments $E_1$, $E_2$ and $E_3$, producing an Erlang-distributed latent period with mean $1/\sigma$ and reduced variance relative to a single exponential stage. 

Transmission occurs along the weighted contact network. The force of infection acting on individual $i$ at time $t$ is defined as 
\begin{equation*}
    \lambda_i(t)=\tau\sum_{j\in I(t)}D_{ij},
\end{equation*}
where $D_{ij}$ denotes the duration of contact between individuals $i$ and $j$, and $\tau$ is the per unit-duration transmission rate. The parameters $1/\sigma$ and $1/\gamma$ correspond to the mean latent and mean infectious periods respectively. Throughout this study, we fix $\frac{1}{\sigma}=3$ days and $\frac{1}{\gamma}=4$ days, and vary $\tau$ to understand different transmission intensities. 

Contact transmissibility likely depends on additional factors such as frequency, physicality, or location. While the underlying data contains information on these attributes, we assume contact duration is the sole determinant of transmission risk, providing a parsimonious measure with well understood quantification \cite{ferretti2024digital}. This simplification allows us to isolate the structural impact of the reconstructed networks and serves as a proof of concept; more detailed parameterisation is left for future work.

Outbreaks are seeded with a single initial infected individual, chosen randomly with probability proportional to their total weighted degree,
\begin{equation*}
    \mathbb{P}(i\in I(0)) = \frac{\sum_jD_{ij}}{\sum_i\sum_jD_{ij}}.
\end{equation*}
This choice reflects the assumption that individuals with higher cumulative contact duration are more likely both to acquire infection externally and to introduce it into a previously uninfected population.

We characterise transmission potential using $R_0$, defined as the expected number of secondary infections generated by a typical infected individual during the early phase of the epidemic in an otherwise fully susceptible population. In the network setting, $R_0$ depends jointly on the transmission parameter $\tau$, the infectious period $1/\gamma$, and the spectral properties of the weights contact matrix $\boldsymbol{D}$. 

We define $g_\phi$ as the set of individuals infected in generation $\phi$ of the transmission tree. Thus, $g_1= \{\text{index case}\}$, $g_2$ contains the secondary cases directly infected by the index case, $g_3$ contains those infected by individuals in $g_2$ and so forth. Transmission generations are recorded directly from the Gillespie event history.

In our simulations, we estimate the basic reproduction number using the ratio
\begin{equation*}
    R_0 = \frac{|g_3|}{|g_2|},
\end{equation*}
that is, the average number of tertiary infections generated by individuals in generation two. This choice differs slightly from the classical definition of $R_0$, which assumes a single infected individual introduced into an otherwise fully susceptible population. In our case, by the time generation three occurs both $g_1$ and $g_2$ have already been infected, so strictly
\begin{equation*}
    |g_1|+|g_2|\neq 1.
\end{equation*}
However, this definition mitigates bias introduced by the choice of index case. Because the index case is sampled with probability proportional to weighted degree, it is typically more connected than a randomly chosen individual. Using $|g_3|/|g_2|$ instead of $|g_2|/|g_1|$ therefore reduces upward bias caused by preferentially selecting highly connected initial nodes. Moreover, during the early phase of the outbreak the proportion of removed individuals is negligible,
\begin{equation*}
    \frac{|g_1|+|g_2|}{n} \ll 1,
\end{equation*}
so depletion of susceptibles has minimal impact on transmission potential, and the ratio provides a close approximation to the early-epidemic reproduction number. 

In Fig.~\ref{fig: final sizes for each data}, we attempt to explore transmission regimes corresponding to $R_0\in [0,5]$. For some model–data combinations this range is not attainable. In particular, under the Stochastic Block Model (SBM) in both lockdown periods, the network structure constrains the achievable growth rate such that the average estimated $R_0$ does not exceed approximately 2.5 across repeated simulations.

Each point shown in the plot corresponds to a single network realisation with 48 independent stochastic epidemic simulations, chosen to align with parallel execution across a 48-core machine. Averaging over these replicates substantially reduces between-simulation variance and reveals an effective upper bound on $R_0$ imposed by network structure. This bound can be expressed as

\begin{equation*}
    \max(R_0) = \frac{\sum_{i\in g_2} ||\boldsymbol{k}_i||}{|g_2|}-1,
\end{equation*}
where $||\boldsymbol{k}_i||$ denotes the total degree (across all contact durations) of node $i$. This expression corresponds to the mean degree of individuals infected in generation two, minus the single infector (the index case) to whom they are all connected. It therefore represents the maximal number of new infections they could generate in a fully susceptible neighbourhood, conditional on the realised contact structure. In this way, the achievable $R_0$ is constrained not only by the transmission parameter $\tau$, but also by the distribution of degrees among early-generation cases. 

For a single network, we define the basic reproduction number as
\begin{equation*}
    R_0=\frac{\sum_{\text{iters}} |g_3|}{\sum_{\text{iters}}|g_2|},
\end{equation*}
where \textit{iters} denotes the set of stochastic simulation runs performed on the network. This empirical definition of $R_0$ is motivated by consistency in analyses of homogeneous stochastic epidemic models with the theoretical basic reproduction number.

The corresponding final epidemic size, $R_\infty$, is computed from the number of individuals in the recovered compartment once the epidemic has terminated. To avoid bias from stochastic extinction events and fluctuations near the epidemic threshold, we exclude simulations that die out early. We therefore define

\begin{equation*}
    R_\infty=\begin{cases}
        \frac{1}{|\text{iters}|}\sum_{\text{iters}}\frac{ R(\text{end})}{n}&, \text{ if } R_0 \geq1, \\
        0&, \text{ if } R_0 < 1.
    \end{cases}
\end{equation*}
Here, $R(\text{end})$ denotes the total number of recovered individuals at the end of the simulation and $n$ is the population size.

When analysing the estimated dispersion parameter $k$ for each outbreak model, we choose transmission regimes calibrated to $R_0=1.5$. The dispersion parameter provides a standard epidemiological measure of heterogeneity in secondary transmission.

We estimate $k$ by fitting a negative binomial distribution to the empirical secondary case distribution observed in the simulations. In this parameterisation, the distribution is characterised by its mean $R_0$ and dispersion parameter $k$, with variance
\begin{equation*}
    \text{Var}(X)=R_0+\frac{R_0^2}{k}.
\end{equation*}
Thus, $k$ is inversely related to excess variance. As $k\to\infty$, the variance approaches $R_0$, corresponding to a homogeneous Poisson offspring distribution. When $k=1$, the distribution reduces to a geometric distribution and when $k<1$, transmission is overdispersed, meaning a small proportion of individuals account for a disproportionate number of secondary infections (superspreading). 

Empirical estimates for COVID-19 frequently lie between $0.1$ and $0.7$ \cite{adam2020clustering, wang2021superspreading}, although such estimates depend strongly on the population context, circulating variant, behavioural patterns, and public health interventions in place.




\section*{Data Availablity}
The POLYMOD data and CoMix data are available from \href{https://socialcontactdata.org/}{socialcontactdata.org} in a format that feed directly into our computer code.

\section*{Code Availability}
Code to fit the GMM model, generate the network and perform the stochastic simulations is available at \href{https://github.com/LukeMurrayKearney/ML_for_Contact_Networks_and_Epidemics}{https://github.com/LukeMurrayKearney/ML\_for\_Contact\_Networks\_and\_Epidemics}.

\bigskip





\newpage
\begin{appendices}

\section{Network Construction}
\begin{figure}[H]
    \centering
    \includegraphics[width=\linewidth]{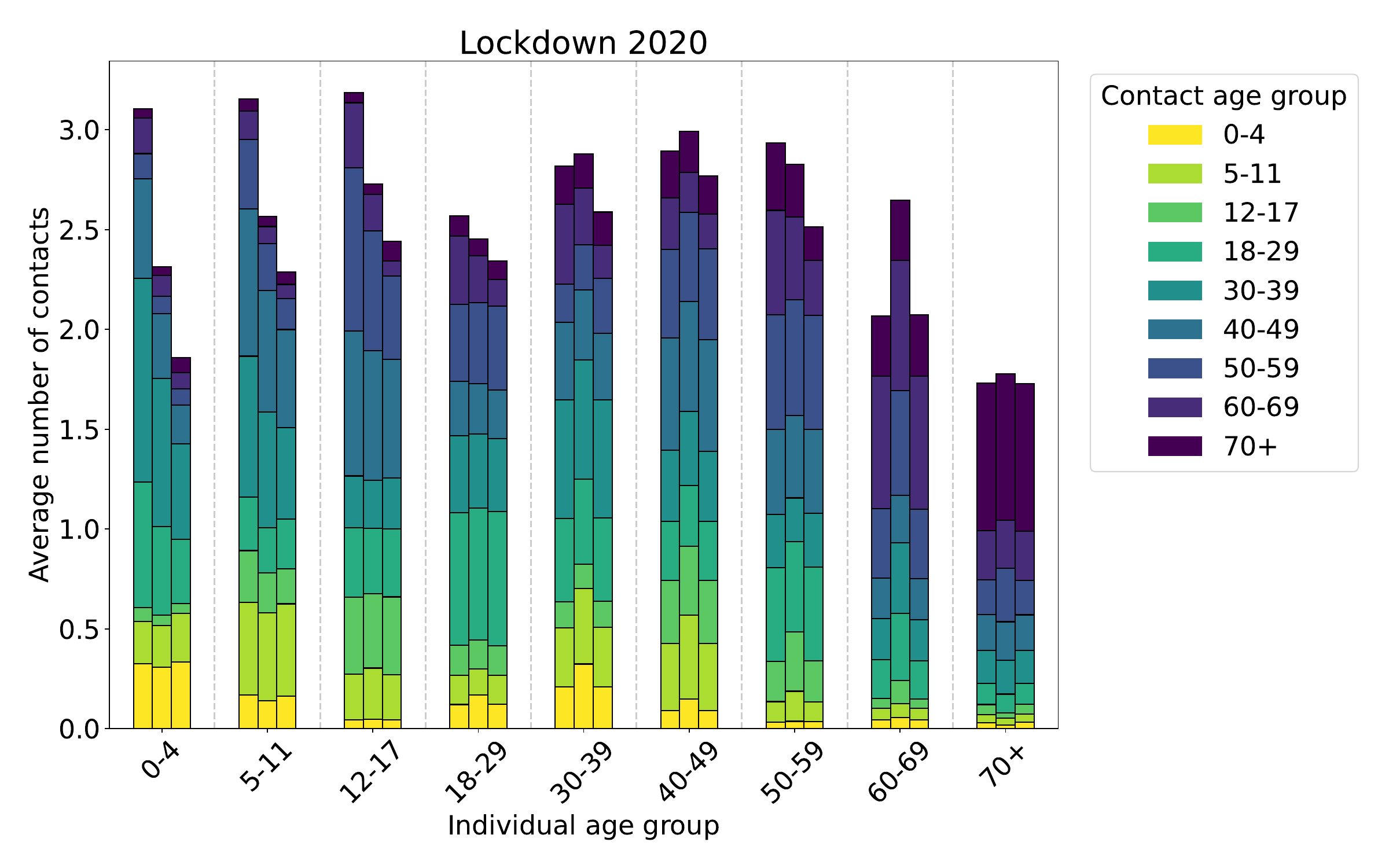}
\end{figure}
\begin{figure}[H]
    \centering
    \includegraphics[width=\linewidth]{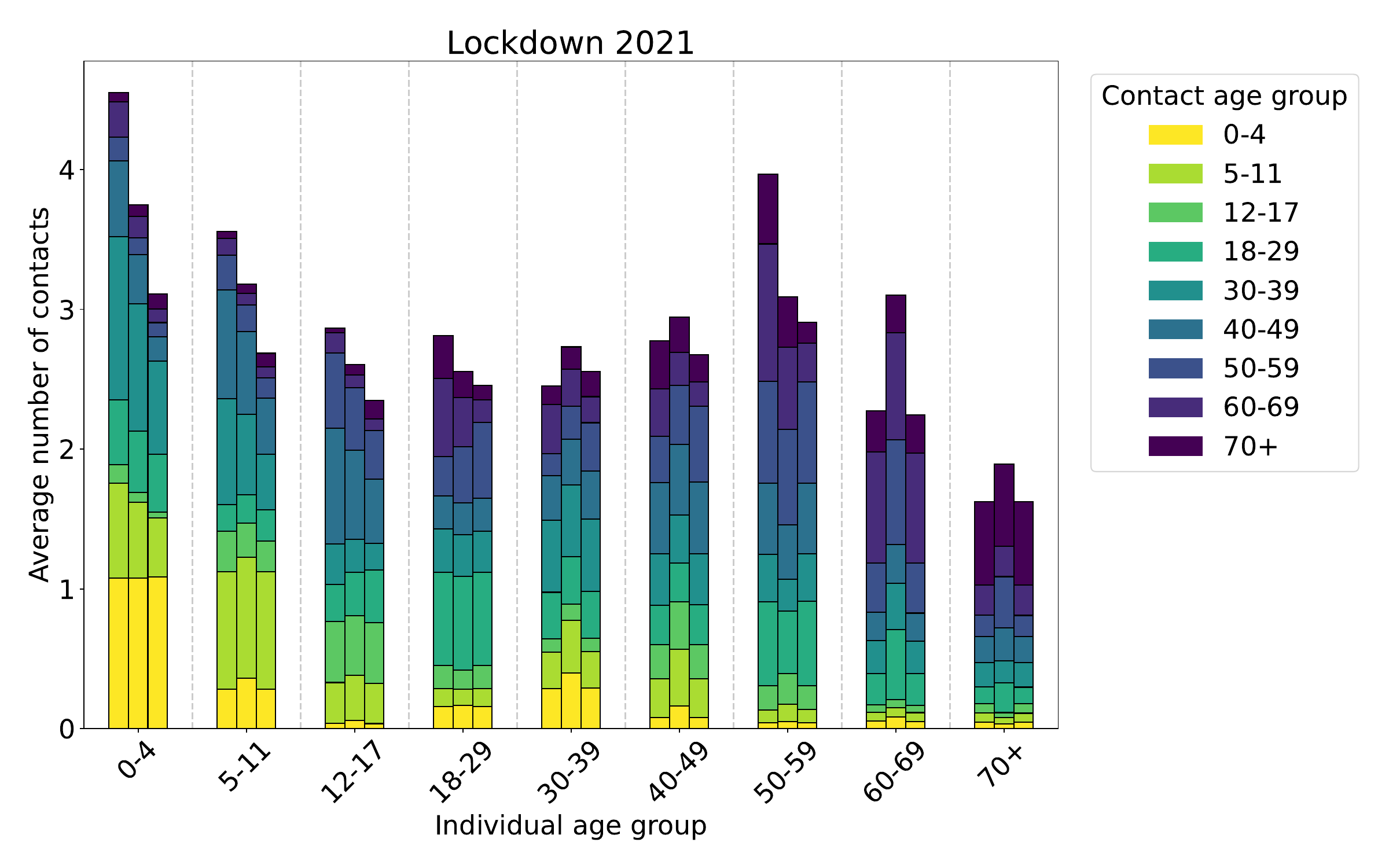}
\end{figure}
\begin{figure}[H]
    \centering
    \includegraphics[width=\linewidth]{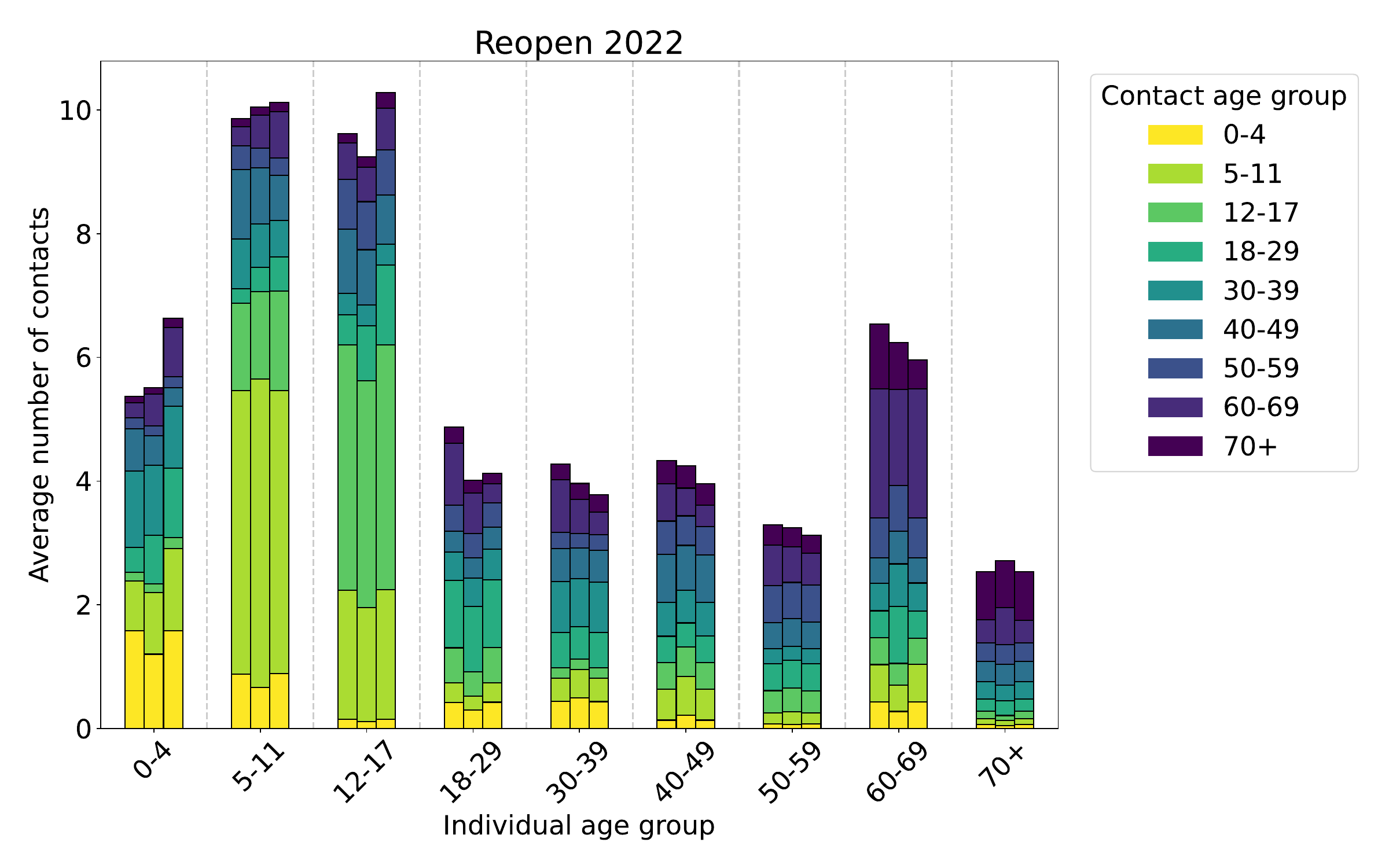}
\end{figure}
\begin{figure}[H]
    \centering
    \includegraphics[width=\linewidth]{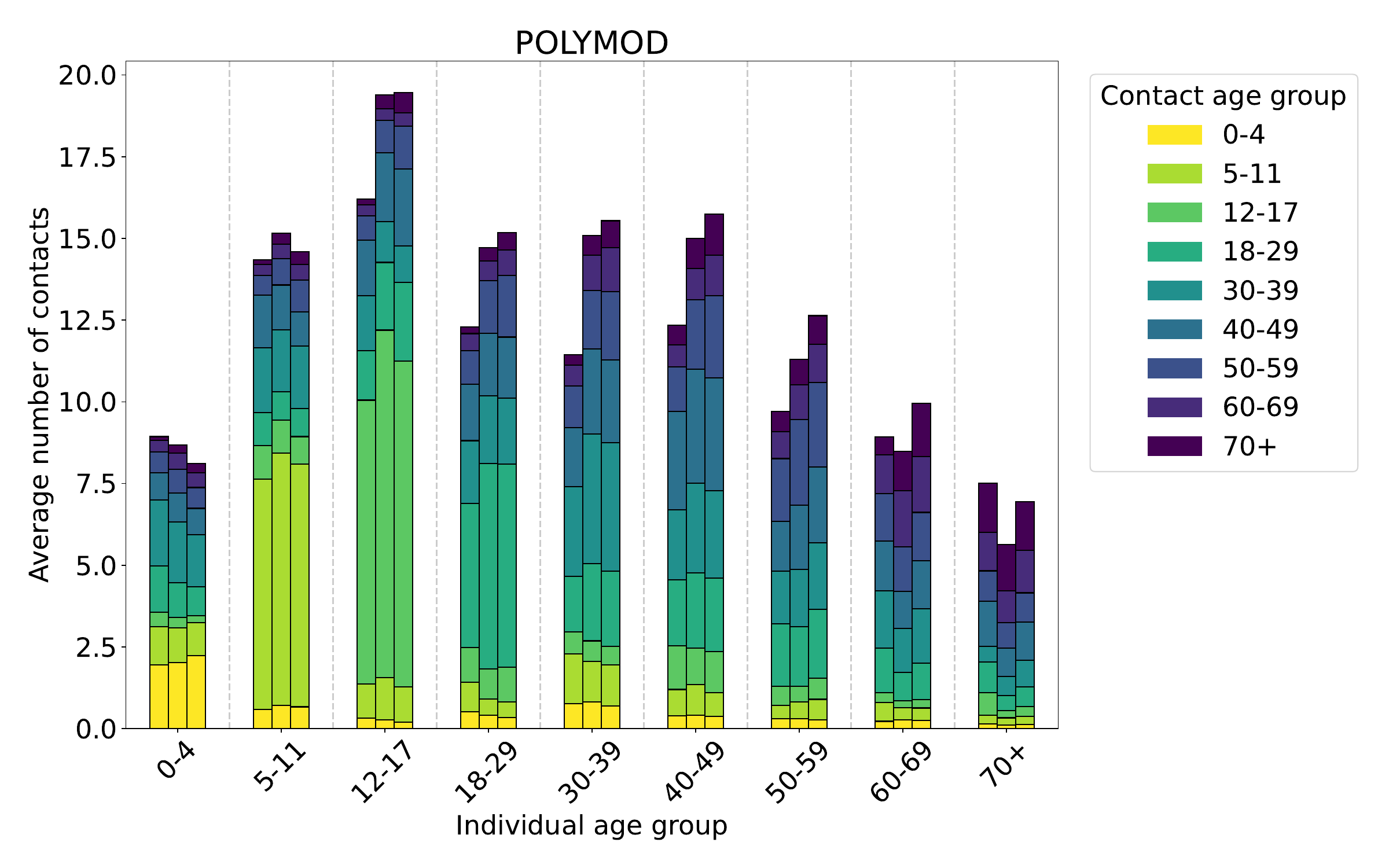}
    \caption{{\bf Average degree distributions} for each age group. The bar triplets refer to the data set, the Gaussian mixture model and the Stochastic block model respectively.}
    \label{fig: Degree distributions}
\end{figure}

\newpage
\section{Network Accuracy}\label{secA1}
\begin{figure}[H]
    \centering
    \includegraphics[width=.74\linewidth]{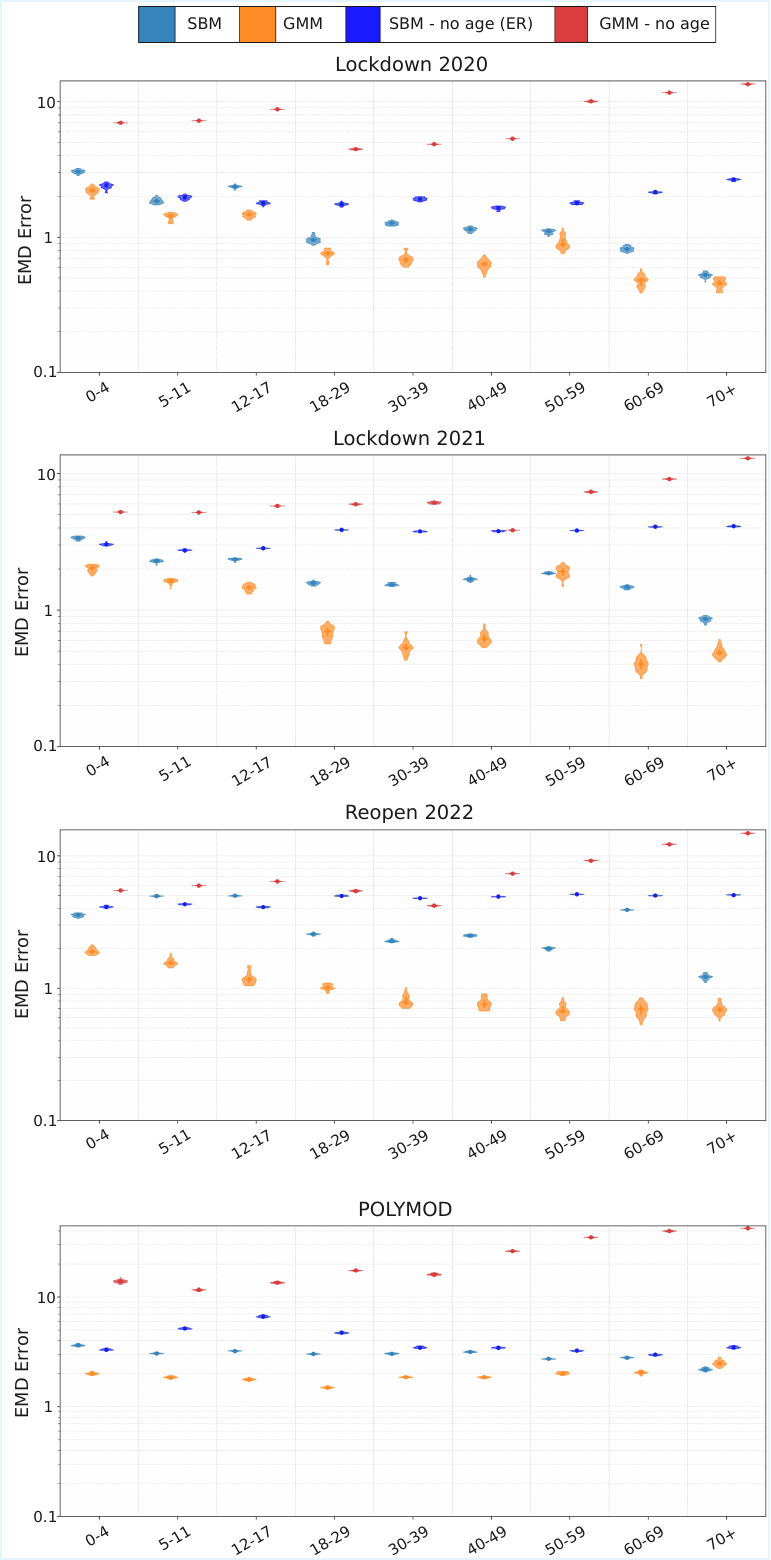}
    \caption{Average Earth Mover's Distance Error (EMD) for each age group and data set.}
    \label{fig:EMD error by age group}
\end{figure}

Repeating the Earth Mover's Distance (EMD) error quantification procedure across multiple independently generated networks and repeated ego-network samples allows us to estimate both the expected reconstruction accuracy and its sampling variability for each network model. Fig.~\ref{fig:EMD error by age group}, shows that the GMM and SBM models with age-structure better represent older individuals in all CoMix data sets, but not in POLYMOD. This correlation between error and age is not found when age-structure is disregarded in network creation.

\newpage
\section{Outbreak Dynamics}

\begin{figure}[H]
    \centering
    \includegraphics[width=.49\linewidth]{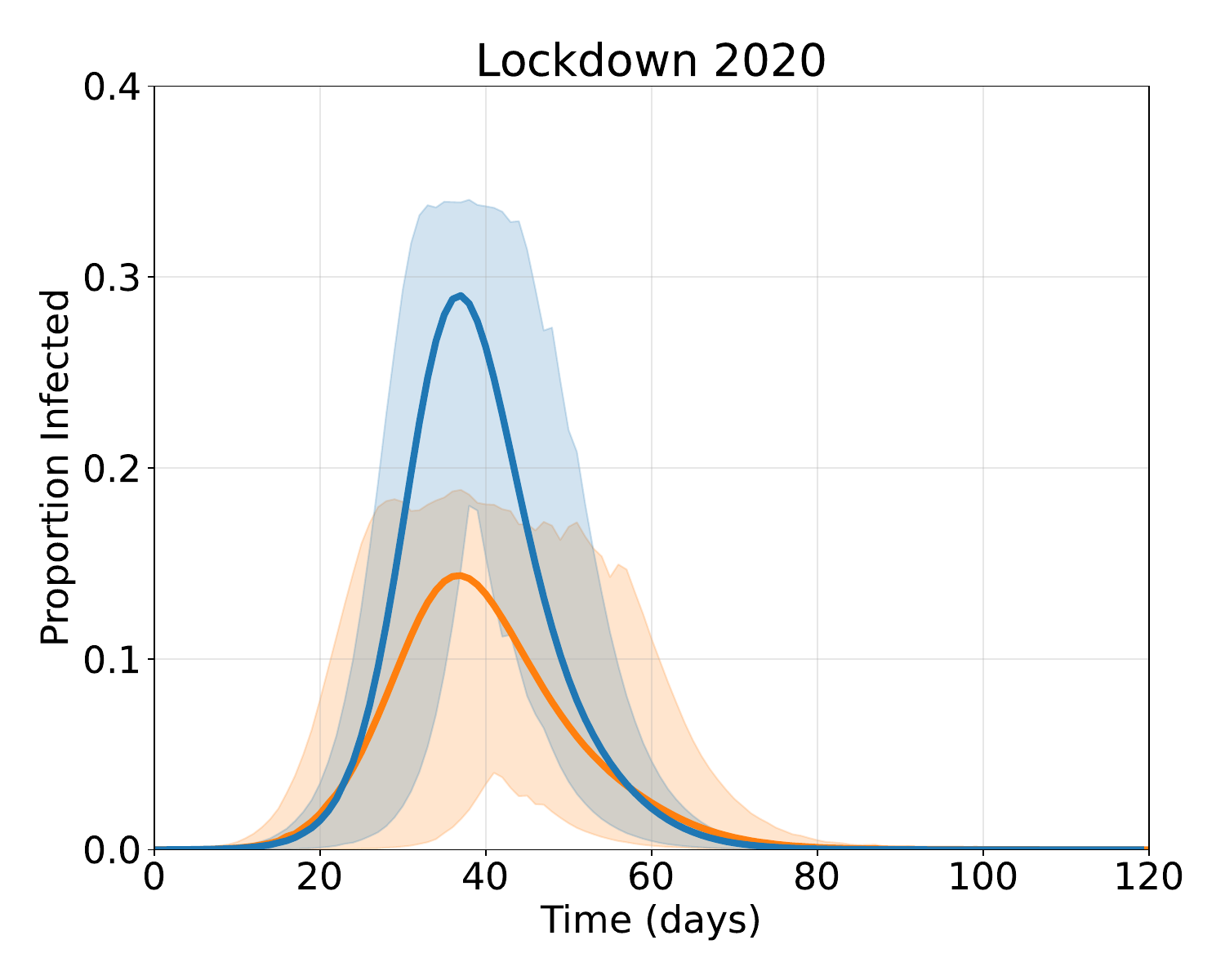}
    \includegraphics[width=0.49\linewidth]{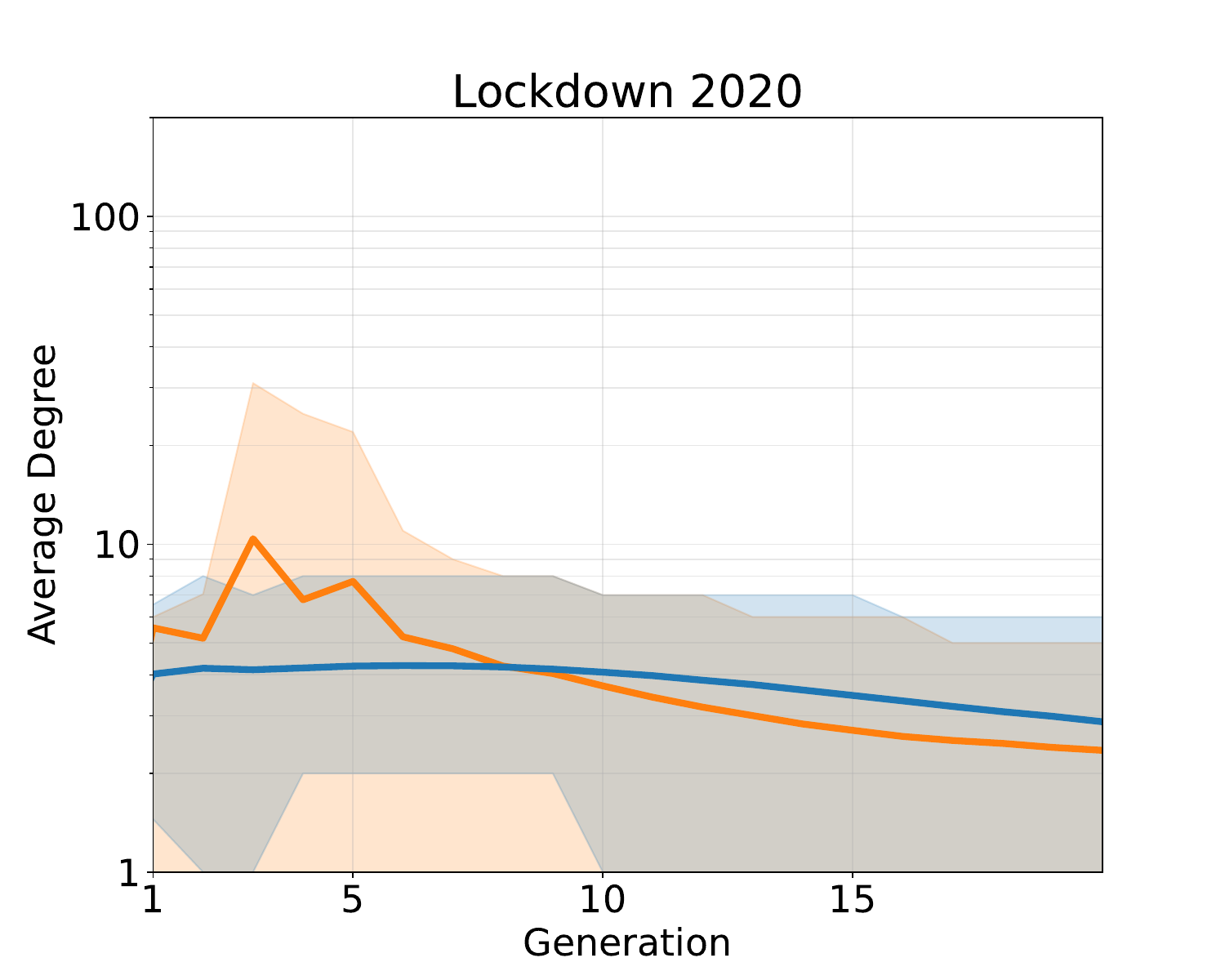}
\end{figure}
\begin{figure}[H]
    \centering
    \includegraphics[width=0.49\linewidth]{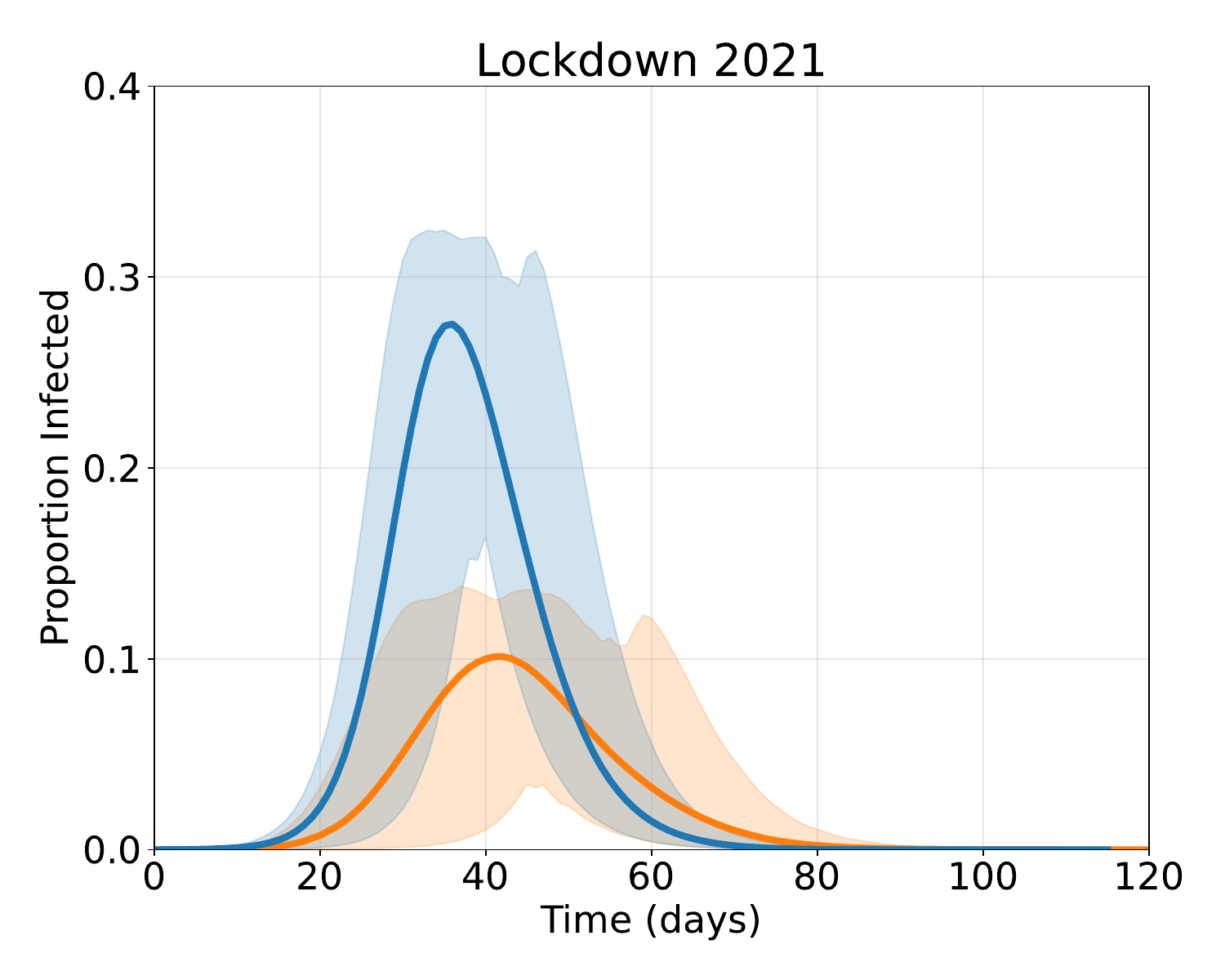}
    \includegraphics[width=.49\linewidth]{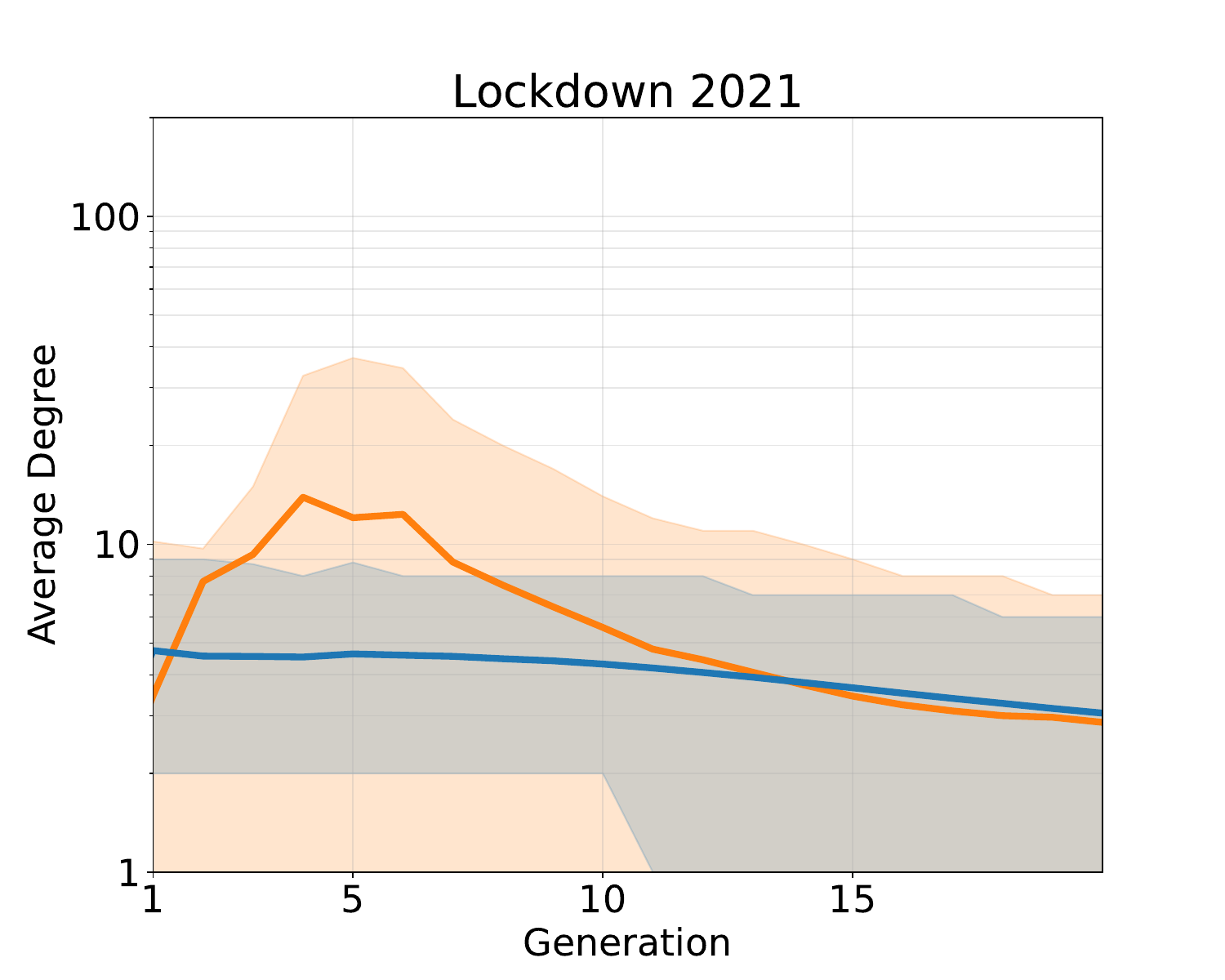}
\end{figure}
\begin{figure}[H]
    \centering
    \includegraphics[width=0.49\linewidth]{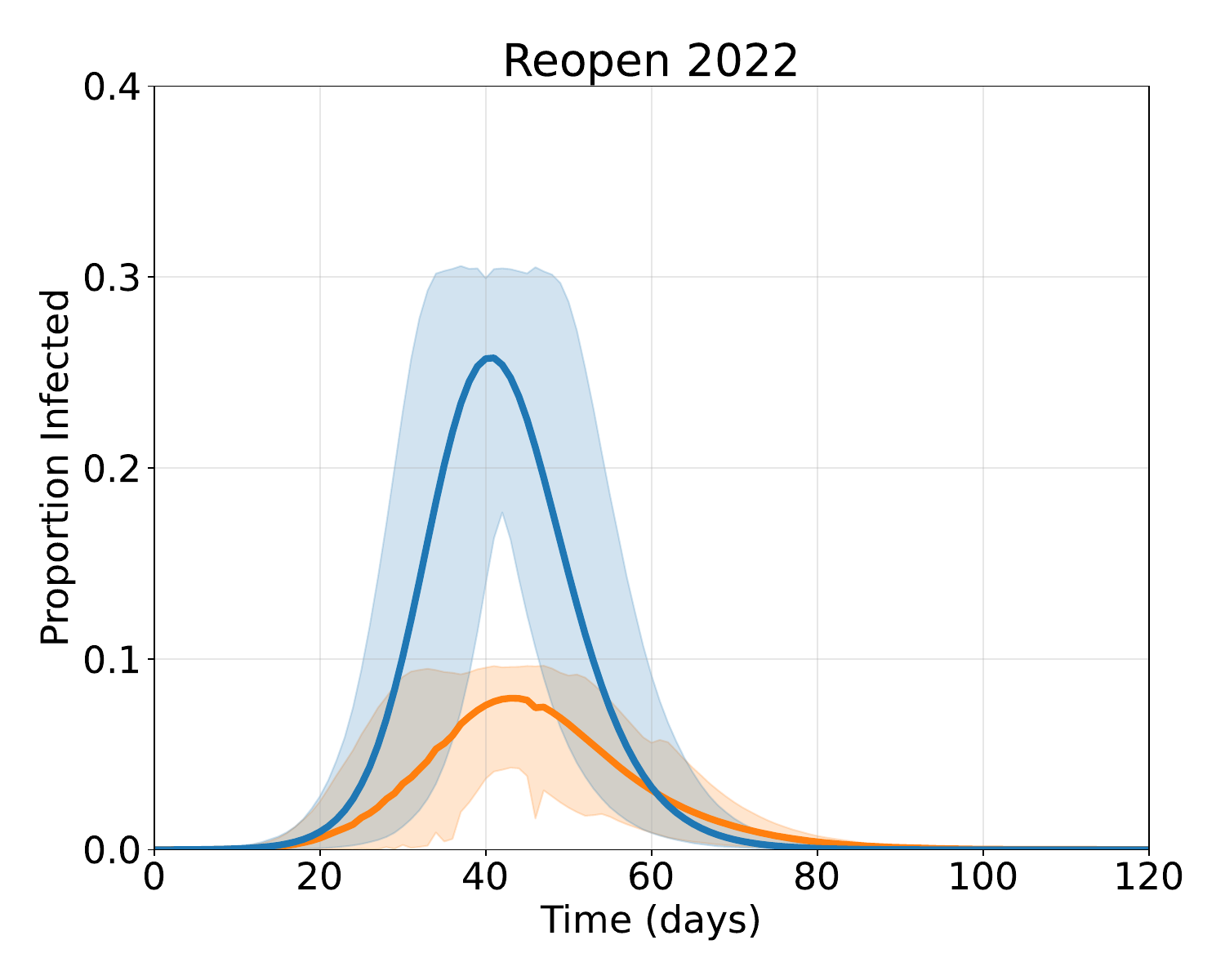}
    \includegraphics[width=.49\linewidth]{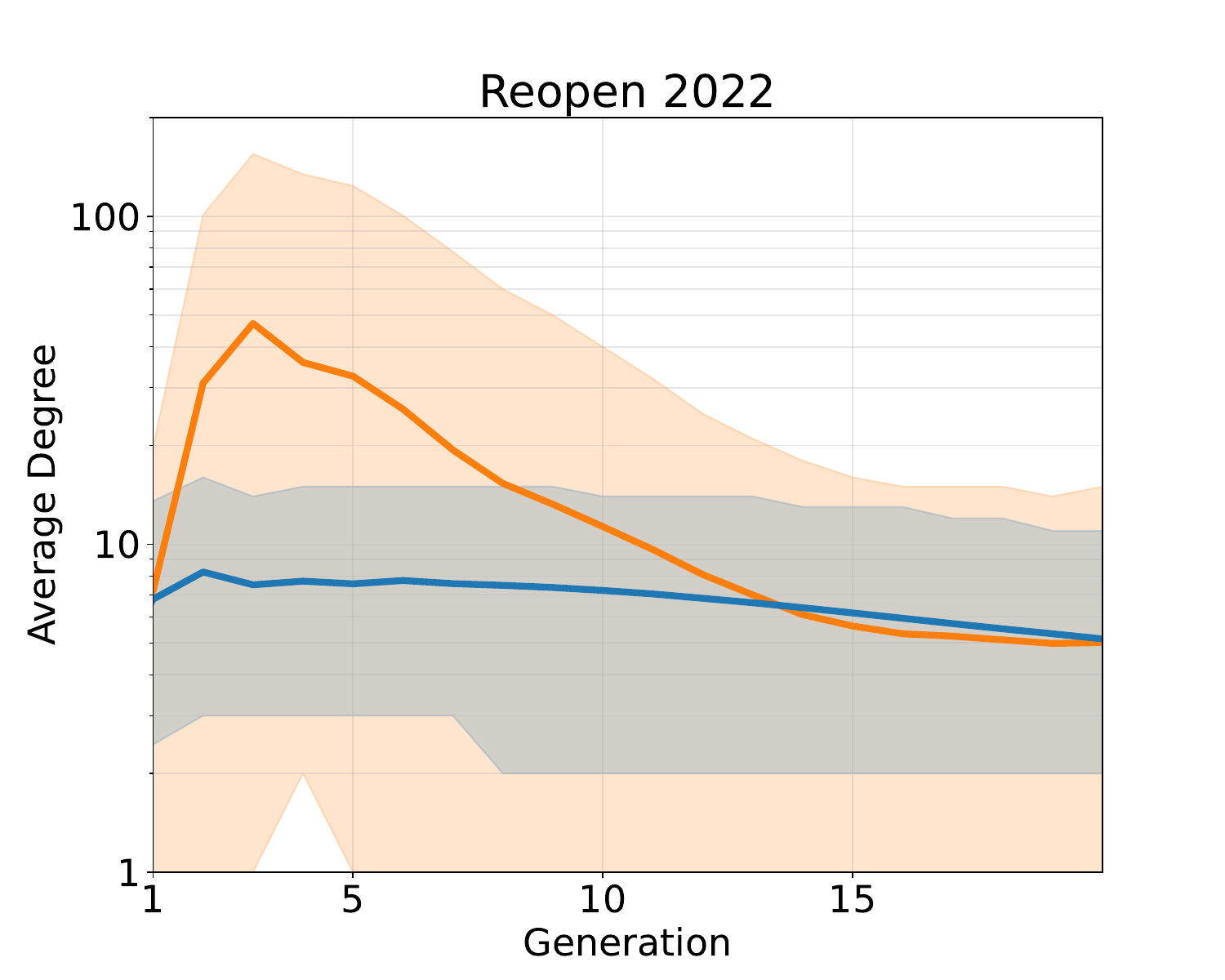}
\end{figure}
\begin{figure}[H]
    \centering
    \includegraphics[width=0.49\linewidth]{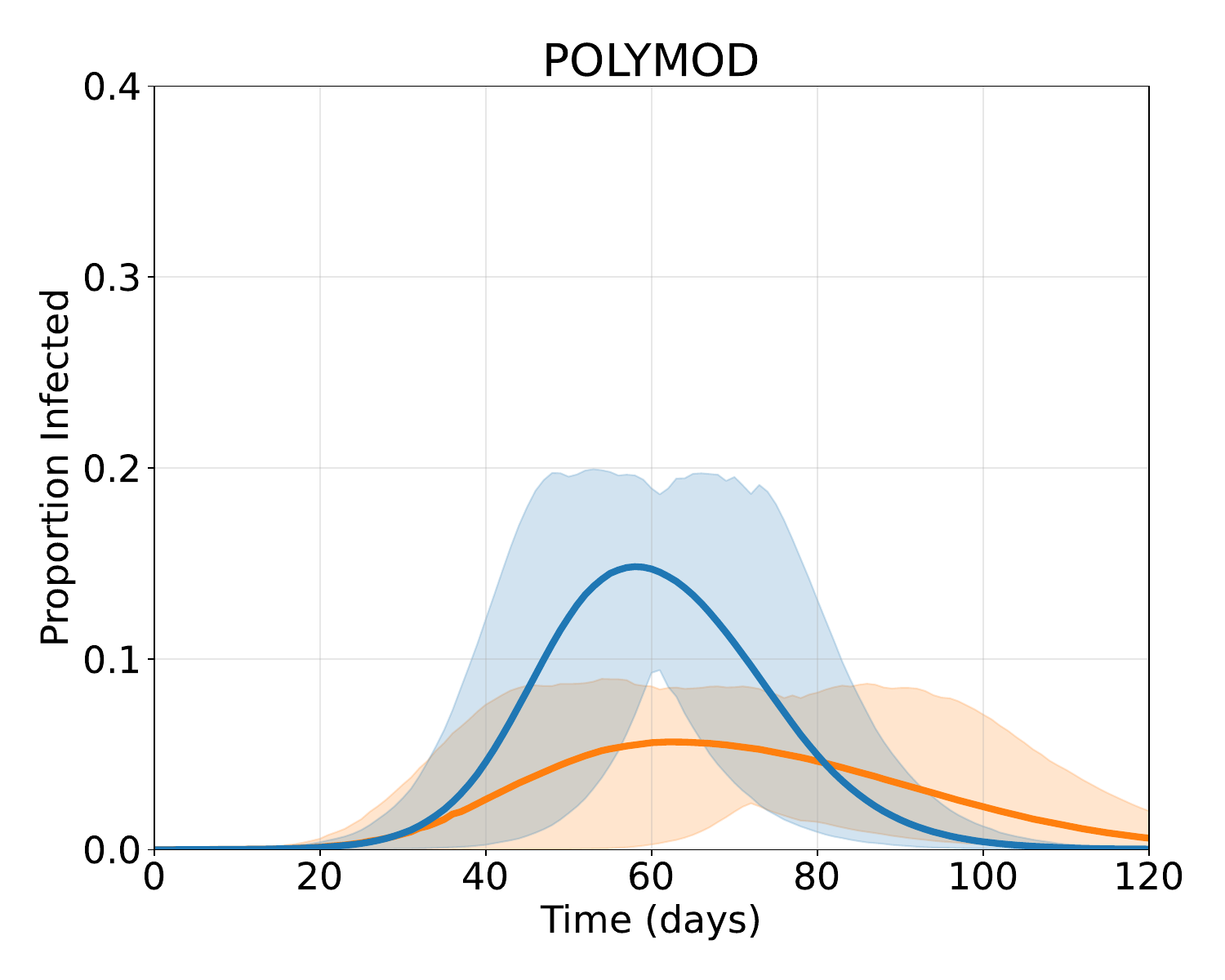}
    \includegraphics[width=.49\linewidth]{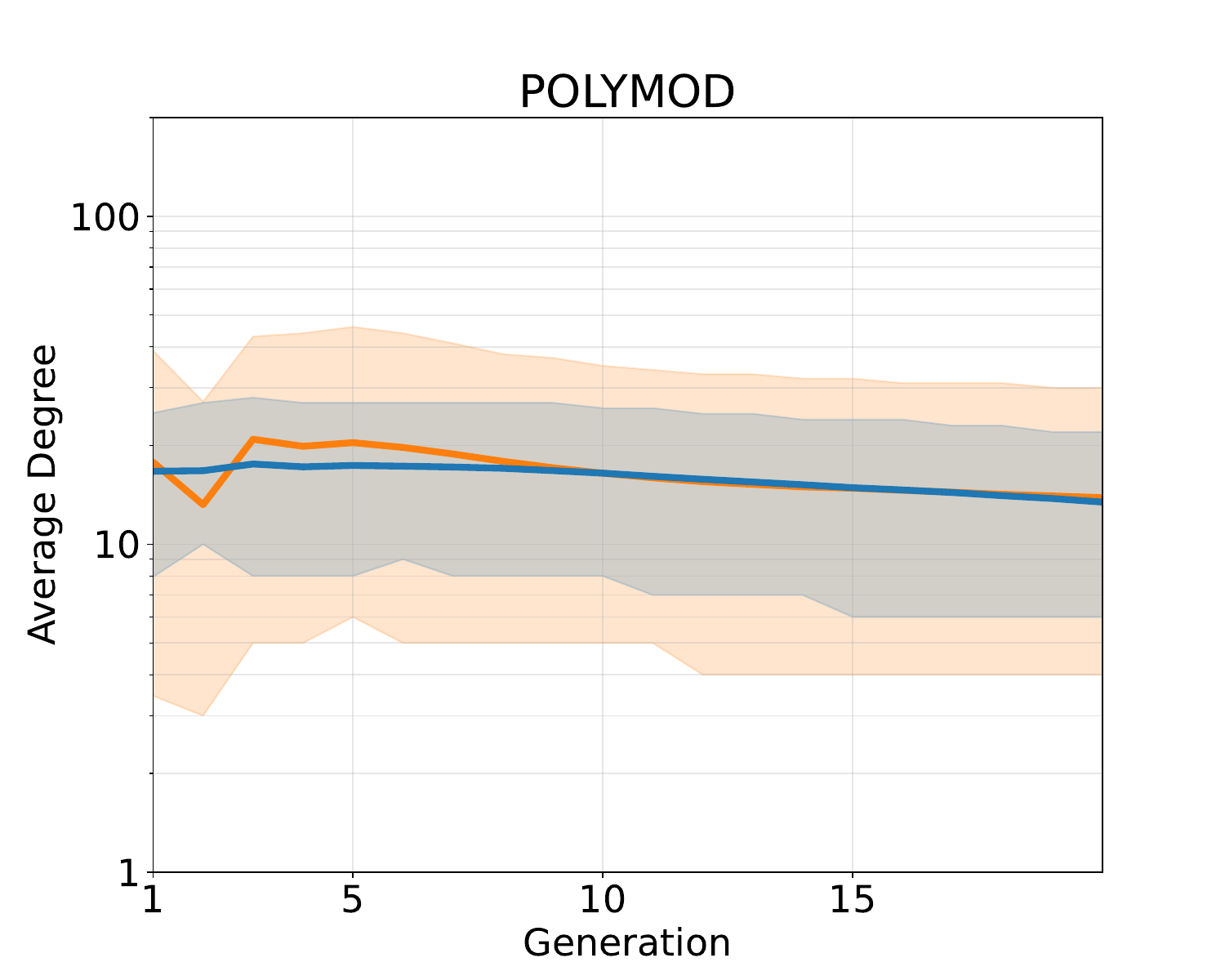}
    \caption{Infection curves and degree change per generation using the GMM and SBM models with duration and an $R_0=2.5$. Averaged over 50 simulations with 95 confidence intervals.}
    \label{fig: Infection curves}
\end{figure}

Fig.~\ref{fig: Infection curves} compares epidemic trajectories and generational changes in mean degree under the GMM and SBM network constructions, both parameterised to yield an early-growth estimate of $R_0=2.5$. Curves are averaged over 50 stochastic simulations, with 95\% confidence intervals shown.

Despite matching early growth rates, the Stochastic Block Model produces substantially larger final epidemic sizes than the GMM-based networks. This divergence arises from structural differences in degree heterogeneity and its interaction with transmission dynamics.

In heterogeneous networks, individuals with higher weighted degree (i.e., larger total contact duration) experience greater force of infection,
\begin{equation*}
    \lambda_i(t)=\tau\sum_{j\in I(t)}D_{ij},
\end{equation*}
and are therefore disproportionately likely to become infected during the early generations of the outbreak. This induces a form of selection on connvectivity, whereby highly connected individuals are removed from the susceptible pool early. In networks with strong degree heterogeneity this process can significantly effect susceptible contact structure over time.

Under the GMM construction, the degree distribution exhibits heavier tails and greater variance. Consequently, early transmission is driven by a small number of highly connected individuals, producing rapid initial growth consistent with $R_0=2.5$. However, once these individuals are infected and removed, the remaining susceptible population is, on average, less connected. The force of infection declines sharply, and epidemic growth slows, often leading to earlier fade-out and smaller final size. 

In contrast, the SBM produces a more homogeneous within-group degree structure. Although it is parameterised to achieve the same early $R_0$, connectivity is more evenly distributed across the population. As a result, depletion of susceptibles does not disproportionately remove highly connected individuals, and the effective reproduction number declines more gradually. This sustains transmission for longer and leads to larger overall epidemic sizes.

These results illustrate that $R_0$ does not uniquely determine epidemic outcomes and is highly dependent on degree heterogeneity. Inducing a disconnect between $R_0$ as an early-growth metric and its classical interpretation as a predictor of epidemic final size under the well-mixed assumption.
\newpage
\subsubsection*{Secondary Case Distribution}
We report the dispersion parameter $k$ of a negative binomial fit to secondary case distribution of outbreaks with $R_0=1.5$ for each model-data pair. We use this framework as a comparative benchmark to assess how well each network model reproduces realistic heterogeneity in transmission. With the exception of the 2021 lockdown period, the Gaussian Mixture Model (GMM) with duration scaling produces $k$ values within the empirically plausible range. When duration weighting is removed, the GMM often becomes excessively overdispersed, yielding $k$ values below the empirical range in approximately half of the study periods. This indicates unrealistically heavy-tailed secondary case distributions driven by extreme connectivity unimpeded by the dampening force of duration.

In contrast, the Stochastic Block Model (SBM), both with and without duration weighting, produces transmission that is too homogeneous. Across all data sets, the estimated dispersion parameter is approximately $k\geq1$, indicating variance close to or below that of a geometric distribution. In these networks, secondary cases are distributed too evenly across individuals, suppressing superspreading behaviour.

These findings demonstrate that reproducing realistic epidemic dynamics requires sufficient structural heterogeneity in the underlying contact network. Models that are too simple underestimate transmission variability, whereas models lacking appropriate duration scaling can over-amplify it. 

It is important to note, however, that the negative binomial assumption itself may be restrictive. If the true secondary case distribution deviates substantially from a negative binomial form, then $k$ may not fully capture transmission heterogeneity. For example, \cite{kremer2021quantifying} show that Poisson log-normal mixture models can provide a better fit to certain empirical COVID-19 transmission data than the negative binomial parameterisation. Consequently, while $k$ offers a convenient and widely used summary statistic, it should be interpreted as an approximate benchmark rather than a complete characterisation of superspreading structure.
\newpage
\subsubsection*{Age and Duration Contributions}
\begin{figure}[H]
    \centering
    \includegraphics[width=\linewidth]{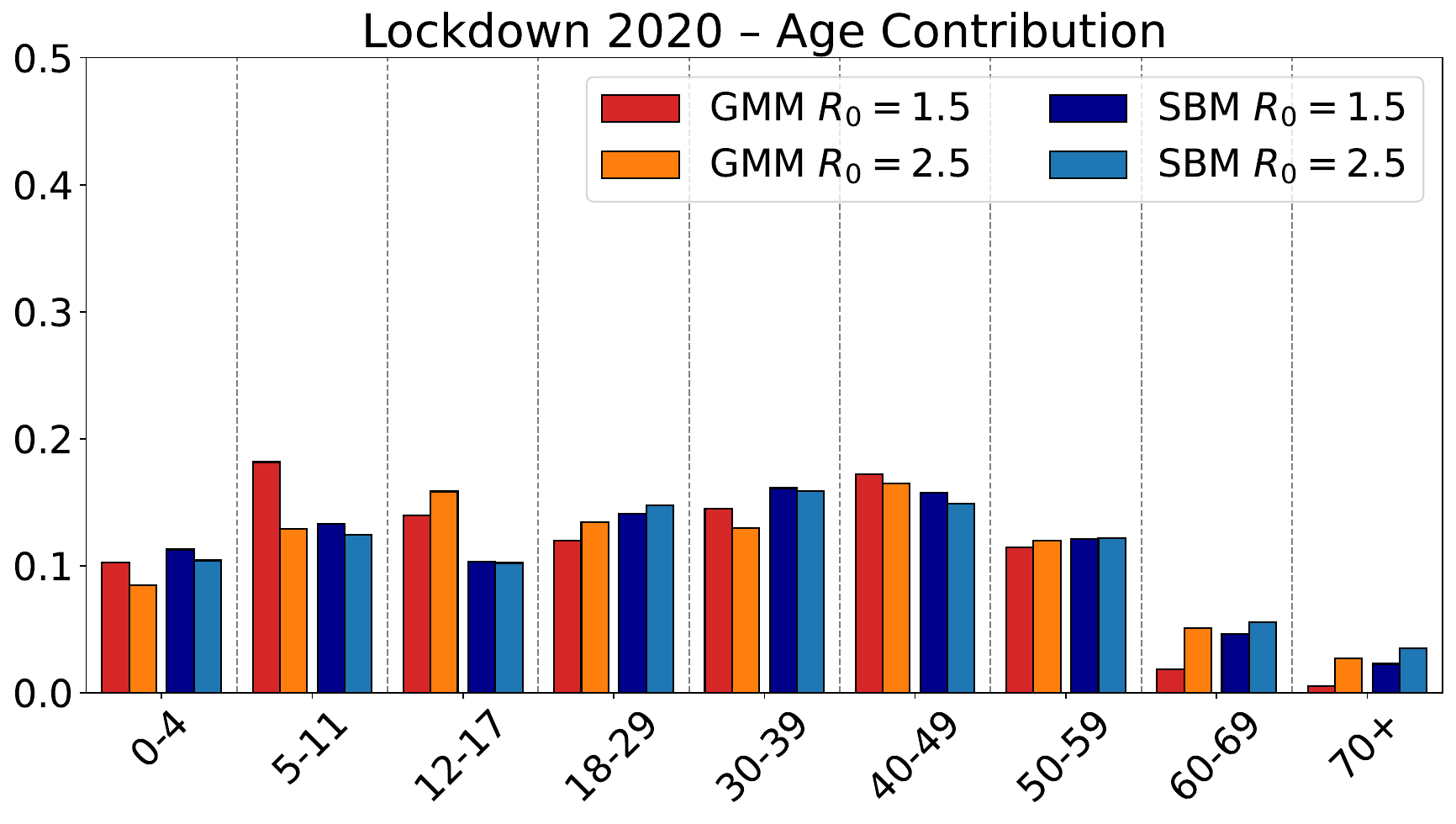}
\end{figure}
\begin{figure}[H]
    \centering
    \includegraphics[width=\linewidth]{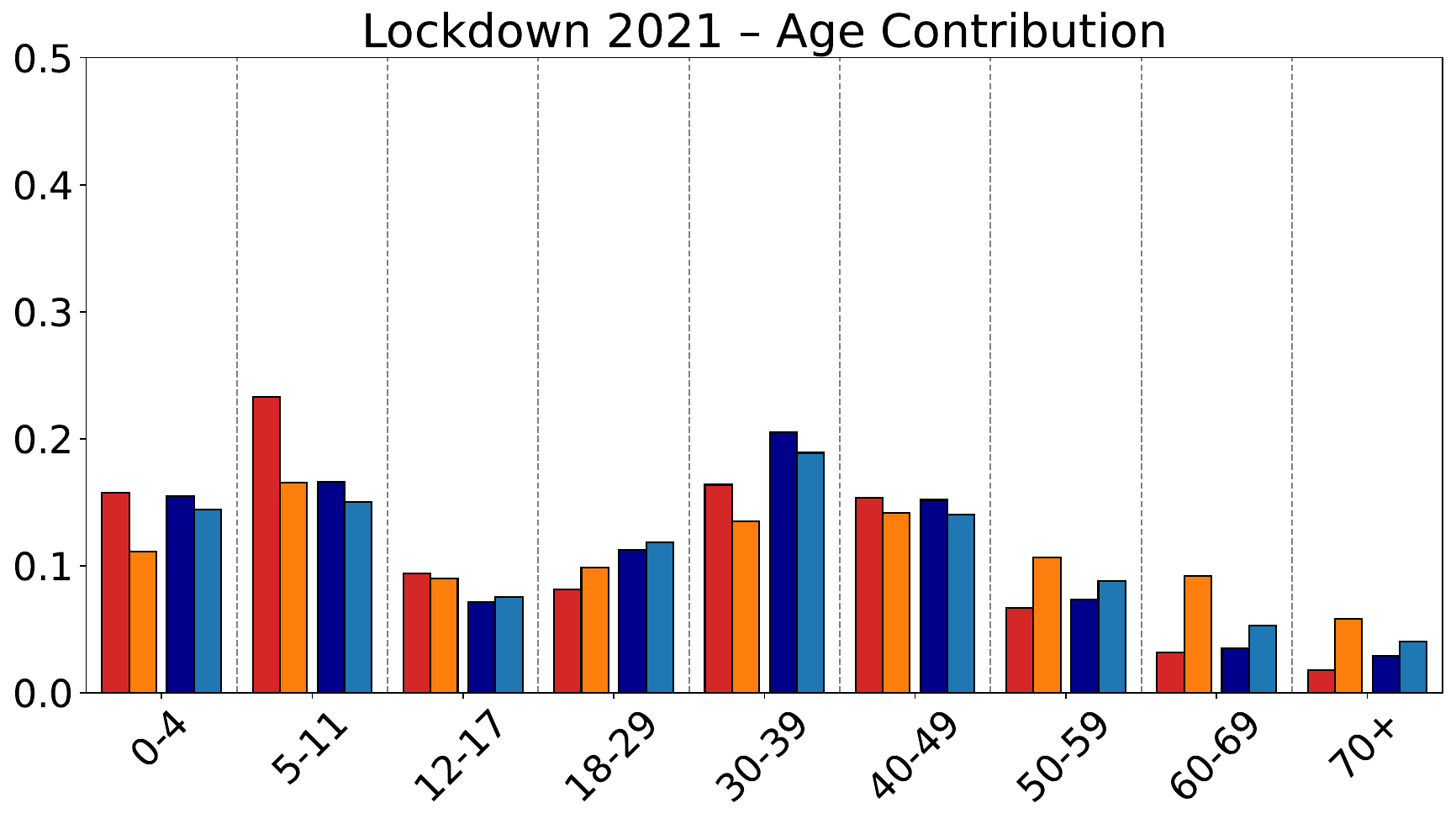}
\end{figure}
\begin{figure}[H]
    \centering
    \includegraphics[width=\linewidth]{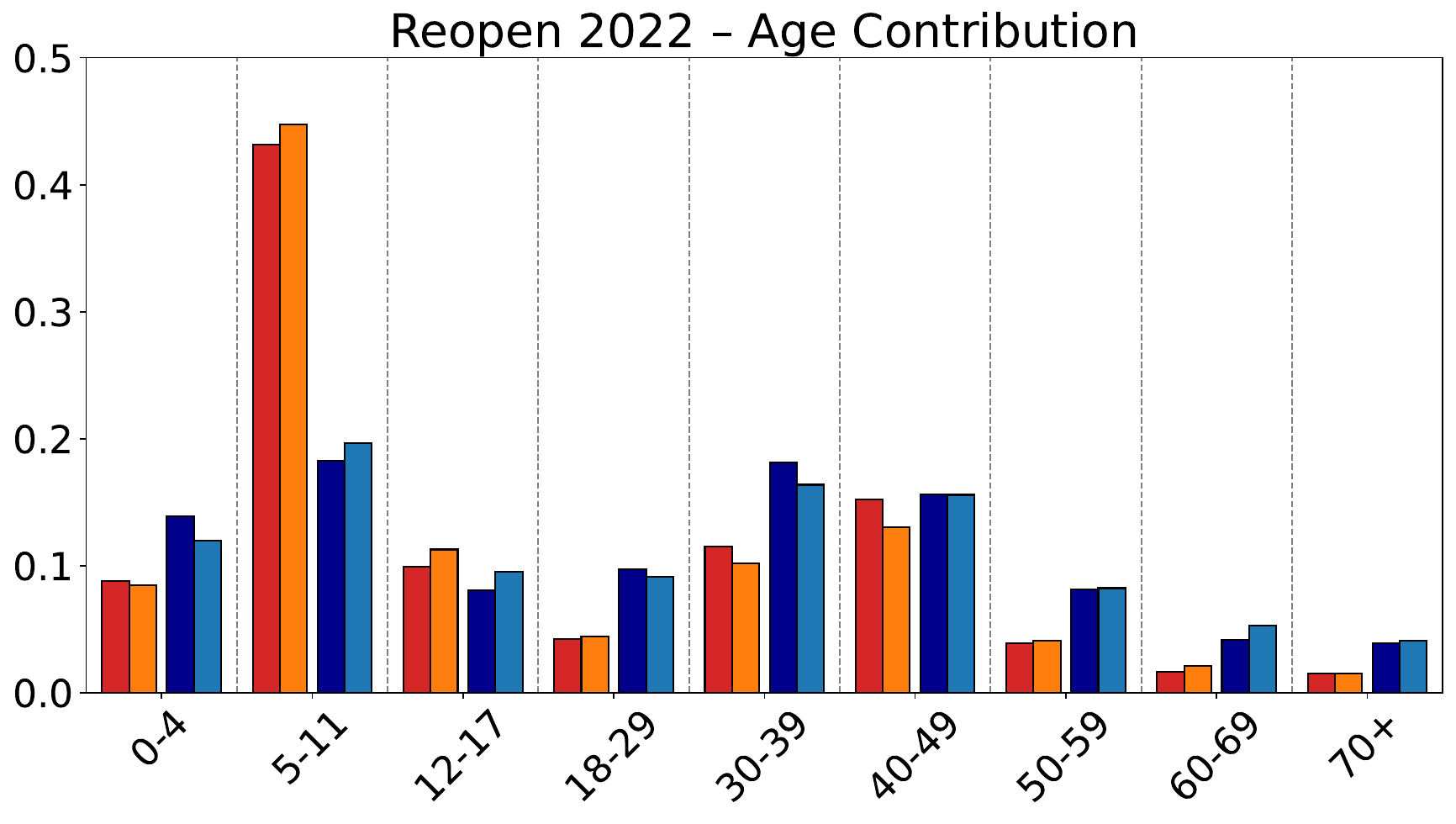}
\end{figure}
\begin{figure}[H]
    \centering
    \includegraphics[width=\linewidth]{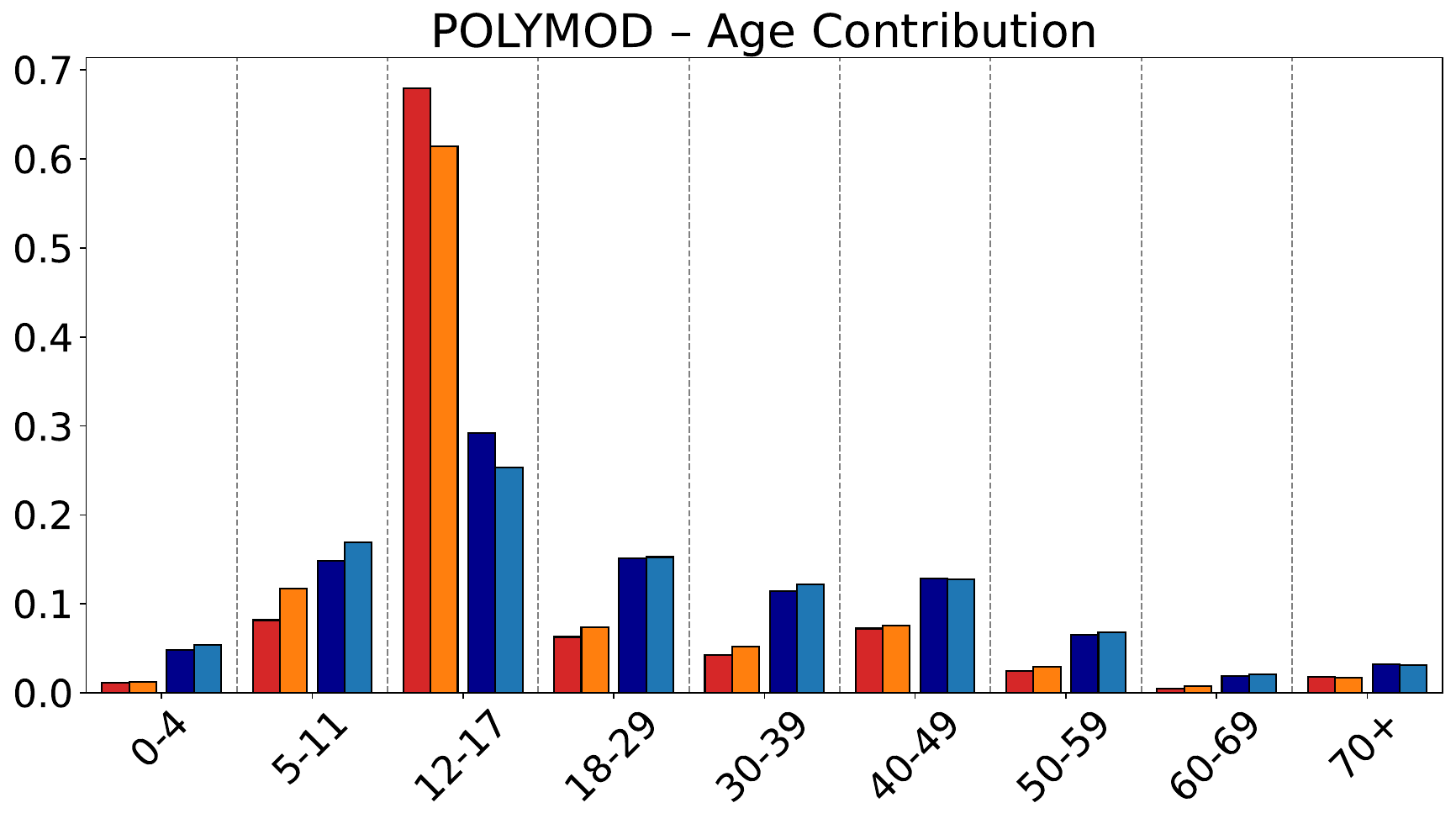}
    \caption{(LHS) Contribution to transmission by differing contact durations. This contribution is quantified by the proportion of infections that are spread through links of each type in the early portion of the outbreak (RHS) Importance of each age-group in transmission. The eigenvector of a $9\times9$ age group matrix created from infection trees in the early portion of the outbreak quantify the importance of each demographic.}
    \label{fig:duration and age contribution supplementary}
\end{figure}

\end{appendices}


\bibliography{sn-bibliography}

\end{document}